\documentclass{emulateapj}

\newcommand{\galex}{GALEX}
\newcommand{\things}{THINGS}

\newcommand{\hi}{\mbox{\rm \ion{H}{1}}}
\newcommand{\hii}{\mbox{\rm \ion{H}{2}}}
\newcommand{\htwo}{\mbox{\rm H$_2$}}

\newcommand{\halpha}{\mbox{\rm H$\alpha$}}

\shorttitle{Extremely Inefficient Star Formation in the Outer Disks of Nearby Galaxies}
\shortauthors{Bigiel et al.}

\begin{document}

\slugcomment{Accepted for Publication in the Astronomical Journal}
\title{Extremely Inefficient Star Formation in the Outer Disks of Nearby Galaxies}

\author{F.~Bigiel\altaffilmark{1,2}, A.~Leroy\altaffilmark{2,3,7},
F.~Walter\altaffilmark{2}, L.~Blitz\altaffilmark{1}, E.~Brinks\altaffilmark{4},
W.~J.~G.~de~Blok\altaffilmark{5}, B.~Madore\altaffilmark{6}}

\altaffiltext{1}{Department of Astronomy, Radio Astronomy Laboratory,
University of California, Berkeley, CA 94720, USA; bigiel@astro.berkeley.edu}
\altaffiltext{2}{Max-Planck-Institut f{\"u}r Astronomie, K{\"o}nigstuhl 17,
69117 Heidelberg, Germany}
\altaffiltext{3}{National Radio Astronomy Observatory, 520 Edgemont Road, Charlottesville,
VA 22903, USA}
\altaffiltext{4}{Centre for Astrophysics Research, University of
  Hertfordshire, Hatfield AL10 9AB, UK}
\altaffiltext{5}{Department of Astronomy, University of Cape Town,
  Rondebosch 7701, South Africa}
\altaffiltext{6}{Observatories of the Carnegie Institution of Washington,
  Pasadena, CA 91101, USA}
\altaffiltext{7} {Hubble Fellow}

\begin{abstract}
We combine data from The \hi\ Nearby Galaxy Survey and the GALEX
Nearby Galaxy Survey to study the relationship between atomic hydrogen
(\hi) and far-ultraviolet (FUV) emission outside the optical radius
($r_{25}$) in 17 spiral and 5 dwarf galaxies. In this regime, \hi\ is
likely to represent most of the ISM and FUV emission to trace recent
star formation with little bias due to extinction, so that the two
quantities closely trace the underlying relationship between gas and
star formation rate (SFR). The azimuthally averaged \hi\ and FUV
intensities both decline with increasing radius in this regime, with
the scale length of the FUV profile typically half that of the
\hi\ profile.  Despite the mismatch in profiles, there is a significant
spatial correlation (at $15\arcsec$ resolution) between local FUV and
\hi\ intensities; near $r_{25}$ this correlation is quite strong, in
fact stronger than anywhere inside $r_{25}$ (where \hi\ is not a good
tracer for the bulk of the ISM), and shows a decline towards larger
radii.  The star formation efficiency (SFE) --- defined as the ratio
of FUV/\hi\ and thus the inverse of the gas depletion time ---
decreases with galactocentric radius across the outer disks, though
much shallower than across the optical disks.  On average, we find the
gas depletion times to be well above a Hubble time ($\sim10^{11}$yr).
We observe a clear relationship between FUV/\hi\ and \hi\ column in
the outer disks, with the SFE increasing with increasing
\hi\ column. Despite observing systematic variations in FUV/\hi, we
find no clear evidence for step-function type star formation
thresholds, though we emphasize that it may not be realistic to expect
them. When compared with results from inside $r_{25}$, we find outer
disk star formation to be distinct in several ways: it is extremely
inefficient (depletion times of many Hubble times which are also long
compared to either the free fall or orbital timescale) with column
densities and SFRs lower than found anywhere inside the optical
disks. It appears that the \hi\ column is one of, perhaps even the key
environmental factor in setting the SFR in outer galaxy disks.
\end{abstract}


\section{Introduction}
When galaxies are observed with sufficient sensitivity, star formation
is often seen to extend well beyond the optical disks, reaching
far into the extended \hi\ disks. After early indications
of extended UV emission in nearby galaxies \citep[e.g.,][]{donas81},
the discovery of extended UV (XUV) disks in a large number of nearby
galaxies was one of the major achievements of the \galex\ mission
\citep{thilker05,gildepaz05,thilker07,gildepaz07b,boissier07,zaritsky07,thilker09,hunter10}.
Extended star formation is seen not only in the UV, but also in the
optical, e.g., in deep \halpha\ and broad brand observations,
which reveal populations of young stars in the outer disks
\citep{ferguson98,lelievre00,cuillandre01,deblok03,christlein08,herbert-fort10,goddard10,werk10}.  This agrees well with the observations that indirect tracers of past or likely future star formation, like
dust \citep{zaritsky94,popescu03,dong08}, CO emission
\citep{braine04,braine07,gardan07}, and metals \citep{gildepaz07b,bresolin09},
are detected in the extended \hi\ envelopes of galaxies.

Star formation at large radii usually does not account for a large
fraction of a galaxy's total star formation rate (SFR), but studying
this process offers a way to illuminate the physics behind the star
formation process {\citep[e.g.,][]{bush08,bush10}: low metallicites and dust abundances, relatively
high shear, low total gas column densities spread over significant scale
heights, a preponderance of \hi\ compared to
\htwo, and a comparatively weak stellar potential well make the ISM in
outer galaxy disks a distinctly different environment compared to the
typical star-forming ISM in the inner part of a galaxy. This contrast
leads to a much lower rate of star formation per unit gas mass at large
galactocentric radii and to the conclusion that the physics behind the conversion of gas into stars
{\em must be} affected by these environmental factors
\citep[e.g.,][]{leroy08}.

A robust, quantitative picture of how the environment in outer disks
affects star formation is important if we want to understand the origins of galaxy
structure. Star formation at large galactocentric radii will affect how chemical
enrichment varies across a galaxy \citep{gildepaz07b} and plays a
critical role in determining the location and form of the break in the exponential
stellar disk \citep{pohlen06}. Also, many galaxies sustain a large
reservoir of (low column density) gas in their outer disks over
evolutionary time scales. Measuring the gas consumption time scale in
this regime for many galaxies and comparing it to the gas consumption
time found for the inner parts of galaxies may provide valuable clues
regarding the role of outer disk gas for fueling star formation over cosmological
times \citep[e.g.,][]{shlosman89,blitz96,bauermeister09}.

In this paper, we study the relationship between atomic gas (\hi) and
star formation at large galactocentric radii. We use state-of-the-art
\hi\ \citep[`The \hi\ Nearby Galaxy Survey', THINGS,][]{walter08} and
UV data \citep[`\galex\ Nearby Galaxy Survey', NGS,][]{gildepaz07a},
which provide the field-of-view and the sensitivity needed to probe
into the outer disks of galaxies while still offering the
resolution to examine the interplay between gas (\hi) and star formation locally,
i.e., on scales of a few times $100$\,pc.

We use these two datasets to study the relationship between \hi\ and
star formation in 22 outer galaxy disks (defined as $r = 1$--$2\times$r$_{25}$). We
assess whether the observed decline of the SFR with galactocentric
radius is predominantly due to a decreasing gas supply and we look for
signs of star formation thresholds, as suggested theoretically and by observations
of sharp truncations in radial distributions of \hii\ regions
\citep[e.g.,][]{martin01}. We examine radial and local variations in the star formation
efficiency (SFE), i.e., the SFR normalized to the \hi\ column (and thus the
inverse of the gas consumption time) and variations in the spatial
correlation of star formation and \hi\ with galactocentric radius. We compare our
results for the outer disks of spirals to dwarf galaxies (which share many
of the same environmental factors) and link them to observations from
within the optical disks of an overlapping set of galaxies
\citep{bigiel08,leroy08}.

\section{Data}
\label{data}

We study 22 galaxies: 17 spiral and 5 dwarf galaxies. This sample is
constructed from the overlap of \things\ \citep{walter08} and targets
of the \galex\ NGS \citep{gildepaz07a} that were observed to similar depth
(integration time of $\gtrsim 1.5$\,ks, corresponding to at least one orbit, which
is the standard integration time for \galex\ NGS targets). Table \ref{table-general}
lists our sample along with adopted distance, inclination, position
angle, optical radius $r_{25}$ and morphology \citep[from][except
  that we adopt $i=20\degr$ in NGC~5194]{walter08}. We correct all
maps for inclination using the angles given in Table
\ref{table-general}.

To allow a rigorous comparison, we degrade all
\hi\ and UV maps to a common resolution of $15\arcsec$ (set by the
\hi\ map with the lowest resolution) by convolving with a circular
Gaussian beam. We have carried out a parallel analysis at a matched physical
resolution of 1\,kpc (set by the largest physical resolution in our sample) and find that for the spirals our results for the two cases differ only
marginally. For the five dwarfs in our sample, working at 1\,kpc resolution
significantly reduces the number of independent measurements (per galaxy and total), constraining
our ability to robustly compare the two approaches due to the limited statistics.
In the following we thus only present the matched angular resolution case. The average
physical resolution in our sample is $\sim600$\,pc, varying between $\sim230$\,pc and $\sim1$\,kpc
across the sample.

We focus on the outer disks of galaxies, which we defined to be
  between 1 and 2$\times$r$_{25}$, where $r_{25}$ is the isophotal radius corresponding
  to 25 B-band magnitudes per square arcsecond. This regime is
  illustrated in Figure \ref{fig1} for one galaxy, NGC~3621. The left
  panel shows the THINGS \hi\ and the right panel the GALEX far UV
  map. The overplotted annuli indicate 1 and 2$\times$r$_{25}$ in the
  plane of the galaxy, respectively.

\subsection{A Note on Terminology}

Throughout this paper we compare 21-cm intensity to far UV
intensity. We will refer to 21-cm intensity and the surface densities
of both \hi\ and `gas' interchangeably because we consider it likely
that opacity effects in the 21-cm line and the contribution of
molecular gas are both small in the regime we study. Similarly, we
measure far UV intensities, I$_{\rm FUV}$, from the GALEX maps. We
convert these intensities into an approximate SFR surface density ($\Sigma_{\rm SFR}$)
using the assumptions discussed below (\S\,\ref{data-units}) and will use the
terms `UV intensity' and `star formation rate surface density'
to mean the same thing ($I_{\rm FUV}$). In both
cases, the reader interested in linking our plots directly to
observables has only to make a linear transformation of any axis using
the equations given in this paper.

When we refer to the conversion of \hi\ into stars we implicitly
assume that the gas becomes molecular first, i.e., \hi\ forms into
\htwo\ which then forms stars. Because \htwo\ in outer
disks is not readily observable, we are forced to consider a `zoomed-out'
version of this process, the conversion of \hi\ into stars with
\htwo\ as an unconstrained intermediate phase. \htwo\ likely constitutes
only a small fraction of the gas mass in outer disks, so the question
of what drives the ISM to form stars in outer disks may still be
robustly addressed using only \hi\ and FUV.

There is no set of wide-field CO maps that extends to $2\,r_{25}$
\citep[the widest-field maps reach $\sim r_{25}$,][]{leroy09}. However,
we can readily see that from the (averaged) star formation rate surface densities
that we infer for outer disks ($\Sigma_{\rm SFR} \approx
10^{-6}$--$10^{-4}\,{\rm M}_\odot~{\rm yr}^{-1}~{\rm kpc}^{-2}$, see \S\,\ref{results})
that CO is likely to be very
faint and \htwo\ is only a relatively minor part of the ISM. If these
$\Sigma_{\rm SFR}$ were found in the disk of a spiral galaxy, the
corresponding H$_2$ surface densities would be $\Sigma_{\rm H2} \sim
0.01$--$1$~M$_\odot$~pc$^{-2}$ \citep{bigiel08}, roughly corresponding
to a single giant molecular cloud (M$_{\rm H2} \approx
10^4$--$10^6$~M$_\odot$) per resolution element. This is low
enough to assume that \hi\ comfortably dominates the ISM across our data 
on (approximately) kpc scales and implies
CO intensities well below the detection limits of most existing maps.
Because of this negligible contribution of H$_2$
to the gas budget and the lack of CO observations, we are
confident using \hi\ emission to trace the bulk distribution of the
mass in the ISM in outer disks on kpc scales.

Readers interested in comparing this paper to other results should
also note that we neglect any contribution from helium or heavier
elements when quoting gas surface densities, but that we do take them
into account when quoting star formation efficiencies or gas depletion times.

\begin{figure*}
\plotone{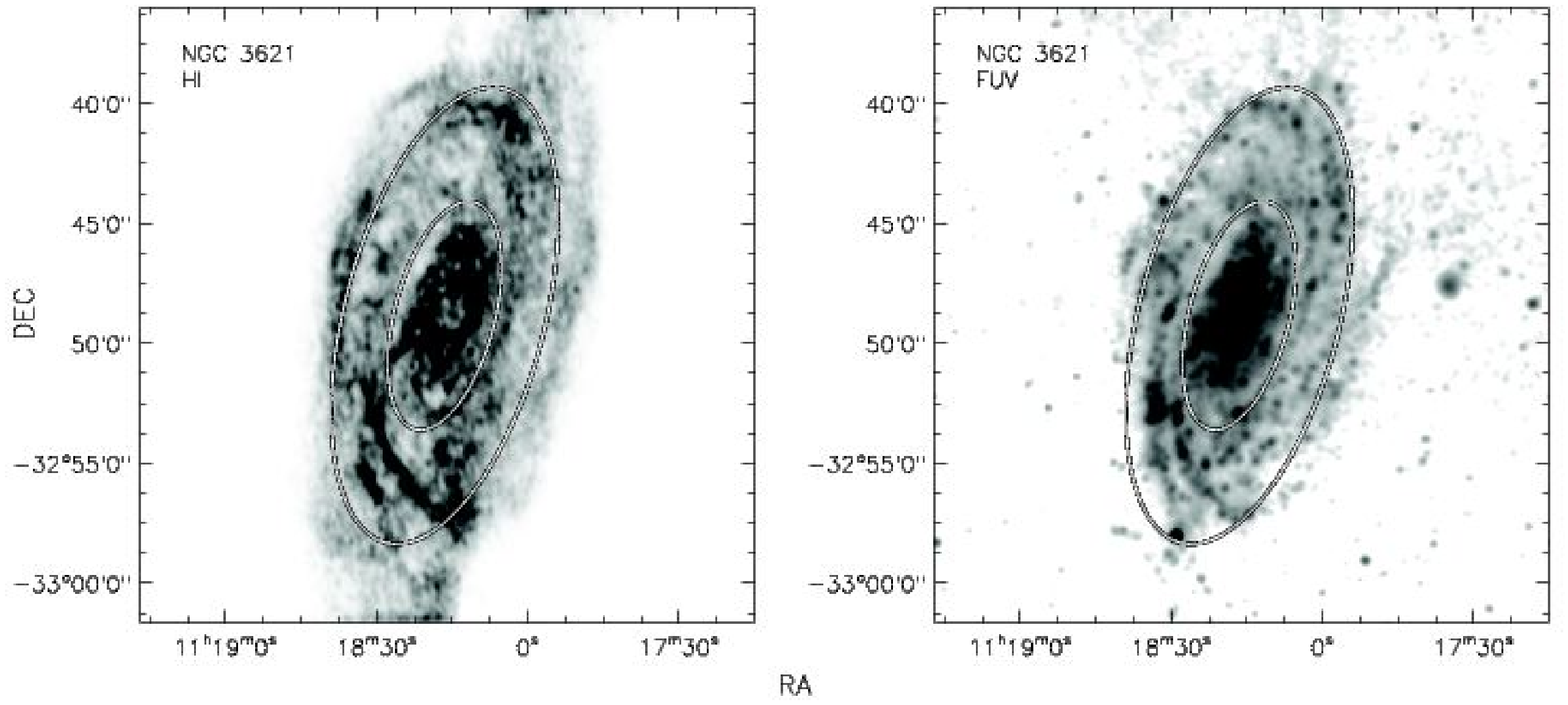}
\caption{THINGS {\sc Hi} (left, linear stretch) and GALEX far UV (right,
  log stretch) maps for NGC~3621. The ellipses indicate 1 and 2
  optical radii (r$_{25}$) in the plane of the galaxy. In this paper,
  we compare \hi\ and far UV emission in this regime for 22 nearby
  galaxies.}
\label{fig1}
\end{figure*}

\begin{deluxetable}{lrrrccc}
\tablecaption{Sample Properties\tablenotemark{1}} \tablehead{
\colhead{Galaxy} & \colhead{D} & \colhead{$i$} & \colhead{PA} & \colhead{$r_{25}$} & \colhead{$r_{25}$} & \colhead{Hubble} \\
\colhead{} & \colhead{[Mpc]} & \colhead{[deg]} & \colhead{[deg]} &
\colhead{[arcmin]} & \colhead{[kpc]} & \colhead{type}} \startdata \tableline
\multicolumn{7}{c}{Dwarfs}\\
\tableline
  DDO\,154 & 4.3 & 66 & 230 & 1.0 & 1.2 & Irr \\
  Ho\,I & 3.8 & 12 & 50 & 1.7 & 1.8 & Irr \\
  Ho\,II & 3.4 & 41 & 177 & 3.3 & 3.3 & Irr \\
  IC\,2574 & 4.0 & 53 & 56 & 6.4 & 7.5 & SABm \\
  NGC\,2366 & 3.4 & 64 & 40 & 2.2 & 2.2 & Irr \\
  \tableline
  \multicolumn{7}{c}{Spirals}\\
  \tableline
  NGC\,0628 & 7.3 & 7 & 20 & 4.9 & 10.4 & Sc \\
  NGC\,0925 & 9.2 & 66 & 287 & 5.4 & 14.3 & Scd \\
  NGC\,2403 & 3.2 & 63 & 124 & 7.9 & 7.4 & SABc \\
  NGC\,2841 & 14.1 & 74 & 153 & 3.5 & 14.2 & Sb \\
  NGC\,2903 & 8.9 & 65 & 204 & 5.9 & 15.2 & SABb \\
  NGC\,3198 & 13.8 & 72 & 215 & 3.2 & 13.0 & Sc \\
  NGC\,3351 & 10.1 & 41 & 192 & 3.6 & 10.6 & Sb \\
  NGC\,3521 & 10.7 & 73 & 340 & 4.2 & 12.9 & SABb \\
  NGC\,3621 & 6.6 & 65 & 345 & 4.9 & 9.4 & SBcd \\
  NGC\,3627 & 9.3 & 62 & 173 & 5.1 & 13.8 & SABb \\
  NGC\,4736 & 4.7 & 41 & 296 & 3.9 & 5.3 & Sab \\
  NGC\,5055 & 10.1 & 59 & 102 & 5.9 & 17.3 & Sbc \\
  NGC\,5194 & 8.0 & 20 & 172 & 3.9 & 9.0 & Sbc \\
  NGC\,5236 & 4.5 & 24 & 225 & 7.7 & 10.1 & Sc \\
  NGC\,5457 & 7.4 & 18 & 39 & 12.0 & 25.8 & SABc \\
  NGC\,7331 & 14.7 & 76 & 168 & 4.6 & 19.5 & Sbc \\
  NGC\,7793 & 3.9 & 50 & 290 & 5.2 & 5.9 & Scd \\
\enddata
\tablenotetext{1}{See \citet{walter08} for further information on individual galaxies and for references to the values quoted in this table.}
\label{table-general}
\end{deluxetable}

\subsection{THINGS \hi}
\label{data-hi}

To estimate the surface density of neutral, atomic hydrogen,
$\Sigma_{\rm HI}$, we use VLA maps of the 21-cm line obtained as part
of `The \hi\ Nearby Galaxy Survey' \citep[THINGS,][]{walter08}. THINGS
consists of high-resolution, high-sensitivity \hi\ data for 34 nearby
galaxies obtained with the NRAO\footnote{The National Radio Astronomy
  Observatory is a facility of the National Science Foundation
  operated under cooperative agreement by Associated Universities,
  Inc.} VLA.  We use `natural' weighted maps, which offer the best
possible signal-to-noise ratios and have an average synthesized beam
size of $\sim\,11\arcsec$. These maps are sensitive ($\sim3\sigma$)
to column densities of $\Sigma_{\rm HI} \gtrsim 0.4~{\rm M}_{\sun}~{\rm pc}^{-2}$
(relatively uniformly across our sample) at our working resolution of
$15\arcsec$. Because the THINGS data include observations in the VLA's
most compact (D) configuration, missing flux is not expected to be a
large concern. For details see \citet{walter08}.

The FWHM of the VLA primary beam (field-of-view) is 32\arcmin .  In
the cases of NGC~5236 (M83) and NGC~5457 (M101), this limits the radii
that we can consider to less than $2\times$r$_{25}$; in these galaxies
the analysis is carried out within $1.7\times$r$_{25}$ and
$1.5\times$r$_{25}$, respectively.

\subsection{\galex\ UV}
\label{data-uv}

To trace recent star formation in the outer disks, we use far UV data from
the `\galex\ Nearby Galaxy Survey'
\citep[NGS,][]{gildepaz07a}. \galex\ provides simultaneous imaging in a
far UV (FUV, 1350\,-\,1750\,\AA) and a near UV (NUV,
1750\,-\,2800\,\AA) broadband filter with angular resolutions (FWHM)
of 4.0\,\arcsec\ and 5.6\,\arcsec\, respectively, and a field-of-view
of 1.25\,\degr\ \citep[for details see][]{morrissey05}. We use only
the FUV band to trace recent star formation, because it is less
sensitive to any old stellar population and suffers less contamination from
foreground stars and background galaxies. We use the NUV band to
identify foreground stars and to estimate the effects of extinction.

We process the NGS maps slightly. First, we identify foreground stars
as regions with an NUV/FUV intensity ratio $> 10$ and a signal-to-noise
ratio $>5$ in the NUV maps. We blank these regions after checking the
results by eye. In a few cases we adjust the color cut-off to a value
higher than 10 to remove particularly bright foreground stars. We also
estimate and remove a small background from the FUV maps, which we
measure away from the galaxy after discarding emission with
intensities more than 3$\sigma$ above the median value of the
image. Finally, we correct the maps for the effects of Galactic
extinction, which we estimate from the extinction maps of \citet{schlegel98}
assuming $A_{FUV} = 8.24 \times E(B-V)$ \citep{wyder07}. In NGC~5194
(M51), we also manually blank the interacting northern companion,
M51b. 

A typical value for the RMS scatter of the noise in our FUV maps at our working resolution is
$\sim2\times10^{-6}\,{\rm mJy\,arcsec^{-2}}$ ($1\sigma$). The exact value
varies significantly from map to map and depends rather sensitively on the
method used to derive it. This estimate reflects the median-based scatter
(which avoids foreground stars and background galaxies) about the zero
level in maps that have already been convolved to $15\arcsec$ resolution.
Variations in exposure time and map quality lead to a factor of 2-3
scatter in the sensitivity of individual galaxies about the value quoted above.

\subsection{Conversion to Physical Units} 
\label{data-units}

We convert 21-cm intensity into atomic gas surface density
via:

\begin{equation} 
\Sigma_{\rm HI}~\left[ {\rm M}_\odot~{\rm pc}^{-2} \right] = 0.015~I_{\rm 21cm}~\left[ {\rm
      K~km~s}^{-1} \right]~,
\end{equation} 

\noindent which yields a hydrogen mass surface density and does not
include heavy elements. To convert FUV intensities into SFR
surface densities, we adopt the conversion of \citet[][their equation
  10]{salim07}. For compatibility with previous work \citep{leroy08, bigiel08}, we divide their
coefficient by an extra factor of 1.59, making the formula appropriate
for a Kroupa-type IMF \citep[compare][Appendix D]{leroy08}. Then

\begin{eqnarray} \label{eq-sfr}
\Sigma_{\rm SFR} \left[ {\rm M}_\odot~{\rm yr}^{-1}~{\rm
    kpc}^{-2}\right] &=& 0.68 \times 10^{-28} \times \\ 
\nonumber
& I_{\rm FUV}& \left[{\rm ergs}~{\rm s}^{-1}~{\rm Hz}^{-1}~{\rm
    kpc}^{-2}\right].
\end{eqnarray} 

\citet{salim07} derived this calibration for $\sim$50,000 galaxies by
fitting population synthesis models to GALEX UV and SDSS optical
multiband photometry. We assume here that the same calibration is
applicable to the FUV intensity averaged over large parts of outer
galaxy disks. A direct check on
this assumption is not straightforward, but \citet{leroy08} compared
various SFR tracers in the inner parts of many of our target galaxies
and found a scatter of $\sim 50\%$, which we will take as our
uncertainty in Equation \ref{eq-sfr}. We derive $\Sigma_{\rm SFR}$
from the \galex\ FUV maps alone, neglecting the effects of internal
extinction. This is both a practical and physical decision. Standard
methods to correct for the effects of internal extinction --- e.g.,
combination with IR intensity, use of NUV/FUV color --- are largely
impossible because of the low signal-to-noise and easy confusion with
background sources in this regime. At the same time, we expect the effects of
extinction to be small, so that introducing a (noisy) correction
factor could easily do more harm than good.

We can estimate a likely upper limit to the bias introduced by
neglecting extinction from our \hi\ maps. In the Galaxy, $N({\rm
  H})/E(B-V)\,=\,5.8\times10^{21}\,{\rm cm}^{-2}\,{\rm mag}^{-1}$
\citep{bohlin78}. This should represent an upper limit to the
extinction for a given \hi\ column in outer disks, which have lower
metallicities and dust abundances than the Milky Way. If we assume that
FUV originates from the midplane (and is thus only affected by half
the dust along a line-of-sight) and that $A_{FUV}/E(B-V) = 8.24$
\citep{wyder07}, then $A_{FUV} = N(HI)\times 7.1\times10^{-22}$ and
the FUV intensity corrected for internal extinction is

\begin{equation}
\label{eq-extinction}
I_{\rm FUV,corr} = I_{\rm FUV} \times 10^{N(HI)\times 2.8\times10^{-22}}.
\end{equation}

For an \hi\ column of $3.0\,{\rm M}_{\sun}\,{\rm pc}^{-2}$ (i.e., $\sim 3.7 \times 10^{20}$\,cm$^{-2}$) , a typical value between
1 and 2~$r_{25}$, Equation \ref{eq-extinction} yields a correction
factor of $\sim\,1.3$. Thus we expect an upper limit to extinction
effects to be $\sim 30\%$, and lower at larger radii, where
\hi\ surface densities drop significantly below $3.0\,{\rm M}_{\sun}\,{\rm pc}^{-2}$ and for
any region with a dust-to-gas ratio lower than the Galactic average (almost
certainly all of our disks).

While we expect that extinction is not a large concern in the outer
disks, several plots will display data from within the optical disks (see,
e.g., the radial profiles in \S\,\ref{radprofs}). In this regime,
$\Sigma_{\rm SFR}$, as defined here, will represent a (significant) underestimate of the
true SFR surface density (due to dust extinction). As emphasized above, the conversion
between $\Sigma_{\rm SFR}$ and the (observed) FUV intensity $I_{\rm FUV}$ is linear and it
is thus straightforward to link any of the results or plots in this paper directly to our observables.

\section{Results}
\label{results}

We employ several approaches to assess the gas-SFR relationship in
outer disks. First, we compare radial profiles of \hi\ and FUV between
$1-2 \times r_{25}$ to test how the overall decline in gas and star
formation relate to one another (\S\,\ref{radprofs}). Then, we measure the strength of the
local correlation between $\Sigma_{\rm SFR}$ and $\Sigma_{\rm HI}$ as
a function of galactocentric radius (\S\,\ref{rankcorr}). We subsequently investigate the gas-SFR
relationship directly, testing how $\Sigma_{\rm SFR}$ and the star
formation efficiency depend on $\Sigma_{\rm HI}$ in various regimes (\S\,\ref{sfe})
and carry out a pixel-by-pixel analysis of the outer disk (\S\,\ref{local}). We assess
the role of large scale gravitational instability as a potential
driver for outer disk star formation (\S\,\ref{q}) and compare our outer disk data
to results for the optical disks of galaxies (\S\,\ref{sflaw-r25}) presented by
\citet{bigiel08}.

\subsection{Radial Profiles and Exponential Scalelengths}
\label{radprofs}

\begin{figure*}
\plotone{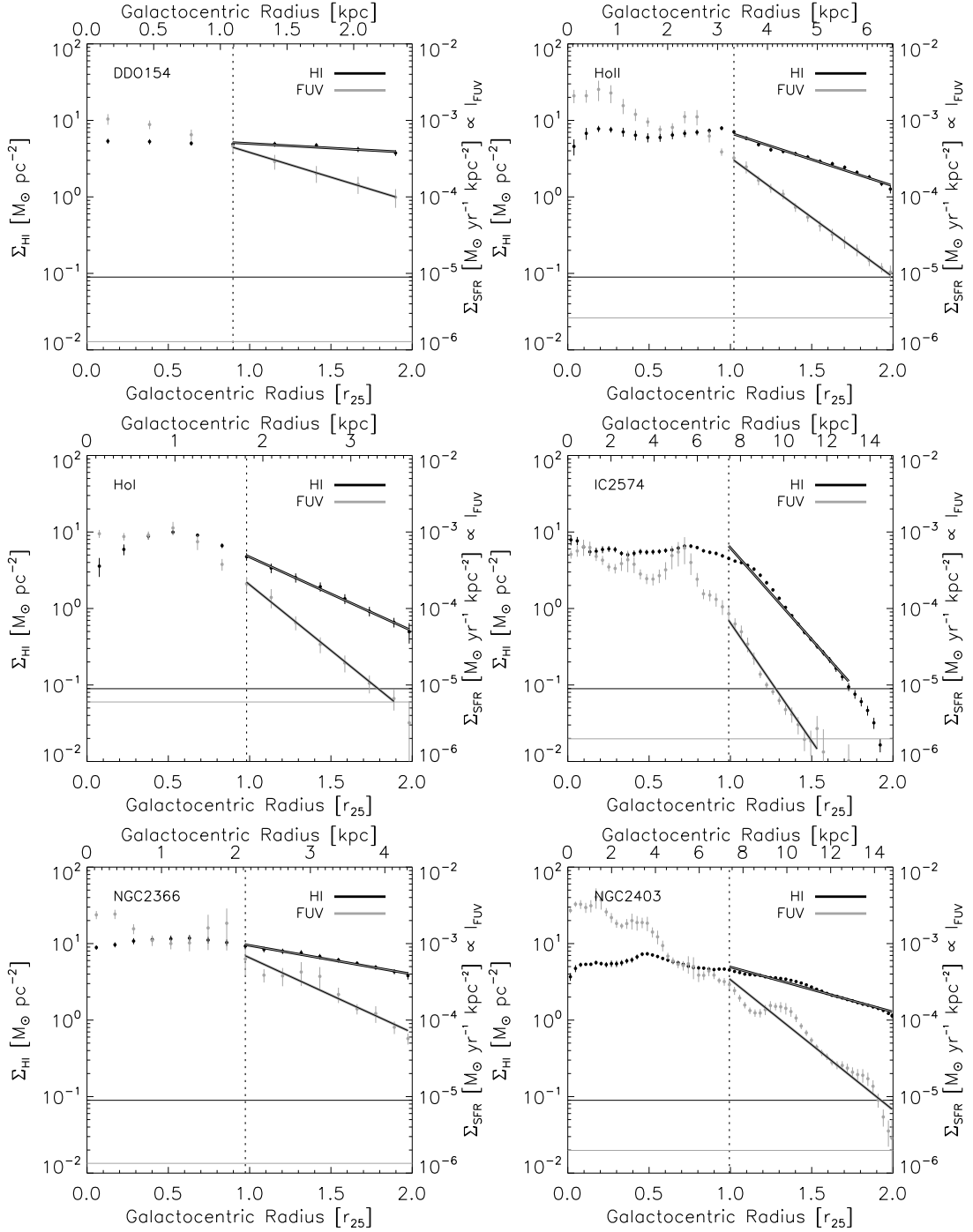}
\caption{Radial profiles of $\Sigma_{\rm SFR}$ (FUV, gray) and
  $\Sigma_{\rm HI}$ (black) for the 22 galaxies in our sample. Error bars
  on each point show the ($1\sigma$) uncertainty in the mean in that ring and
  horizontal lines (black: {\sc Hi}, gray: FUV) show conservative sensitivity estimates based on the 
  line-of-sight sensitivities in our maps and assuming 20 independent measurements per annulus (note
  that the FUV sensitivity is sometimes below the lower plot limit).
  A vertical line indicates the innermost data point included in the fit (approximately
  at $r_{25}$). Solid lines
  show exponential fits to the decline between
  $1-2\times r_{25}$, only considering points above our sensitivity cuts.}
\label{fig2}
\end{figure*}
\begin{figure*}
\figurenum{\ref{fig2}}
\plotone{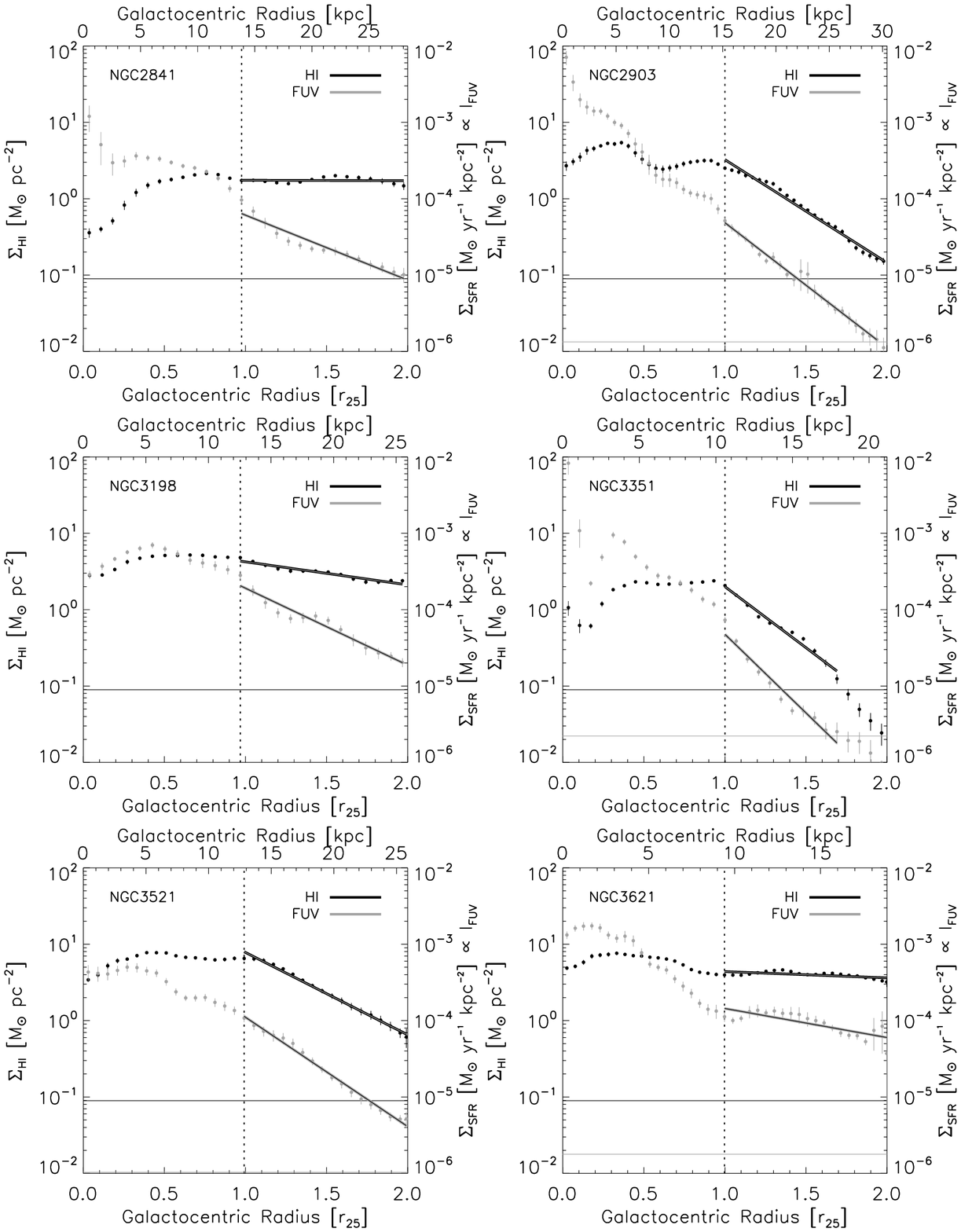}
\caption{continued.}
\end{figure*}
\begin{figure*}
\figurenum{\ref{fig2}}
\plotone{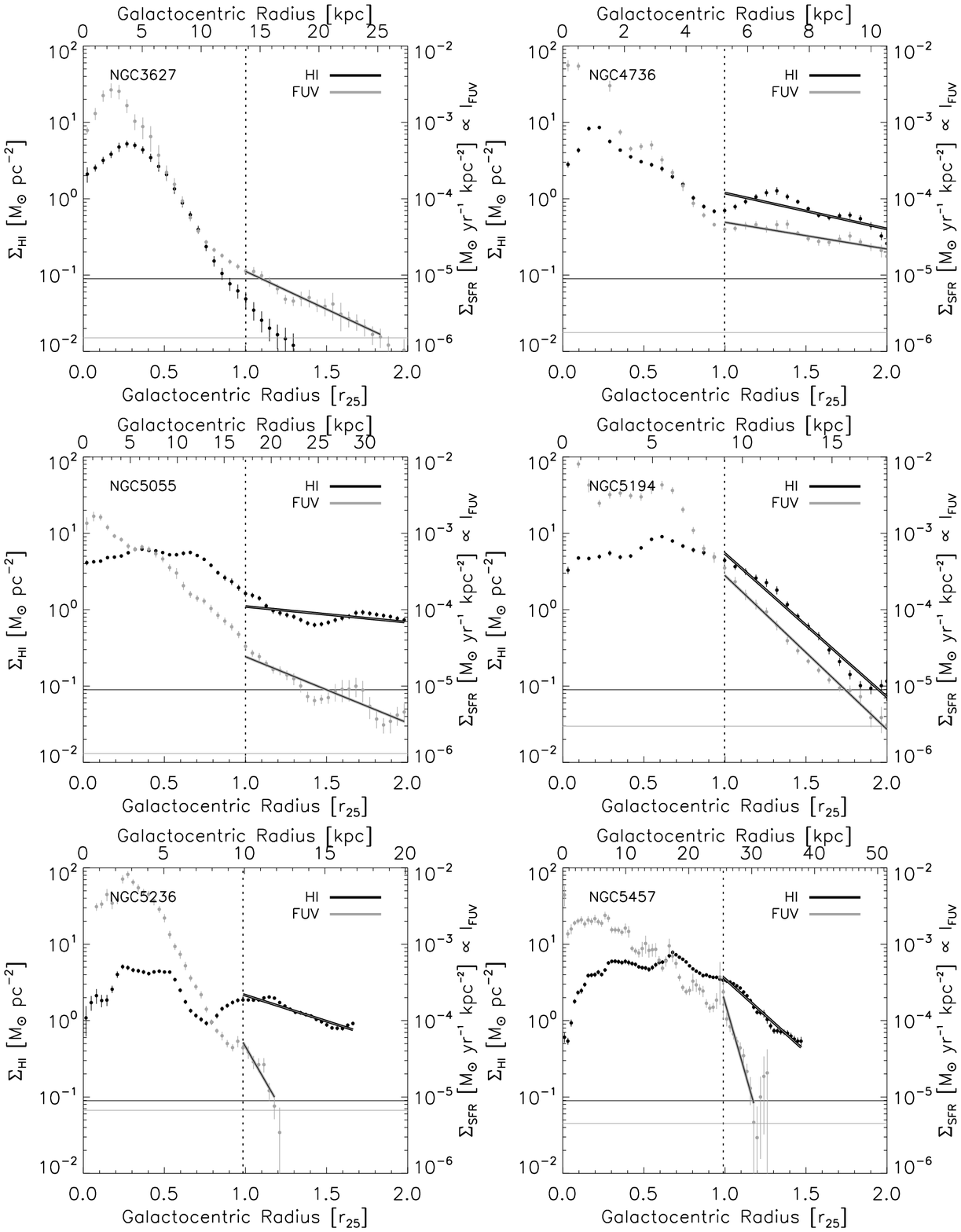}
\caption{continued.}
\end{figure*}
\begin{figure*}
\figurenum{\ref{fig2}}
\plotone{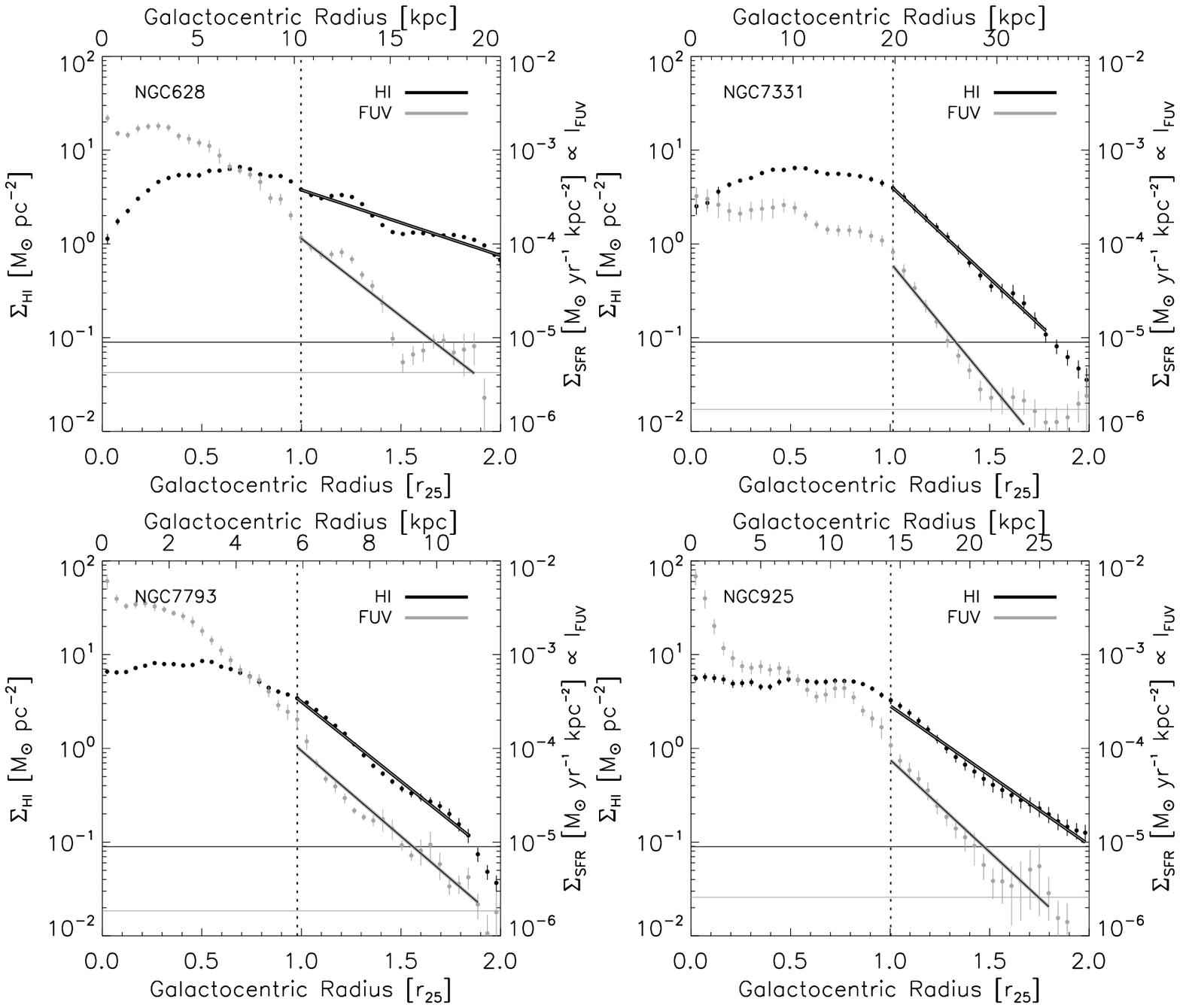}
\caption{continued.}
\end{figure*}

\begin{figure*}
\epsscale{0.6}
\plotone{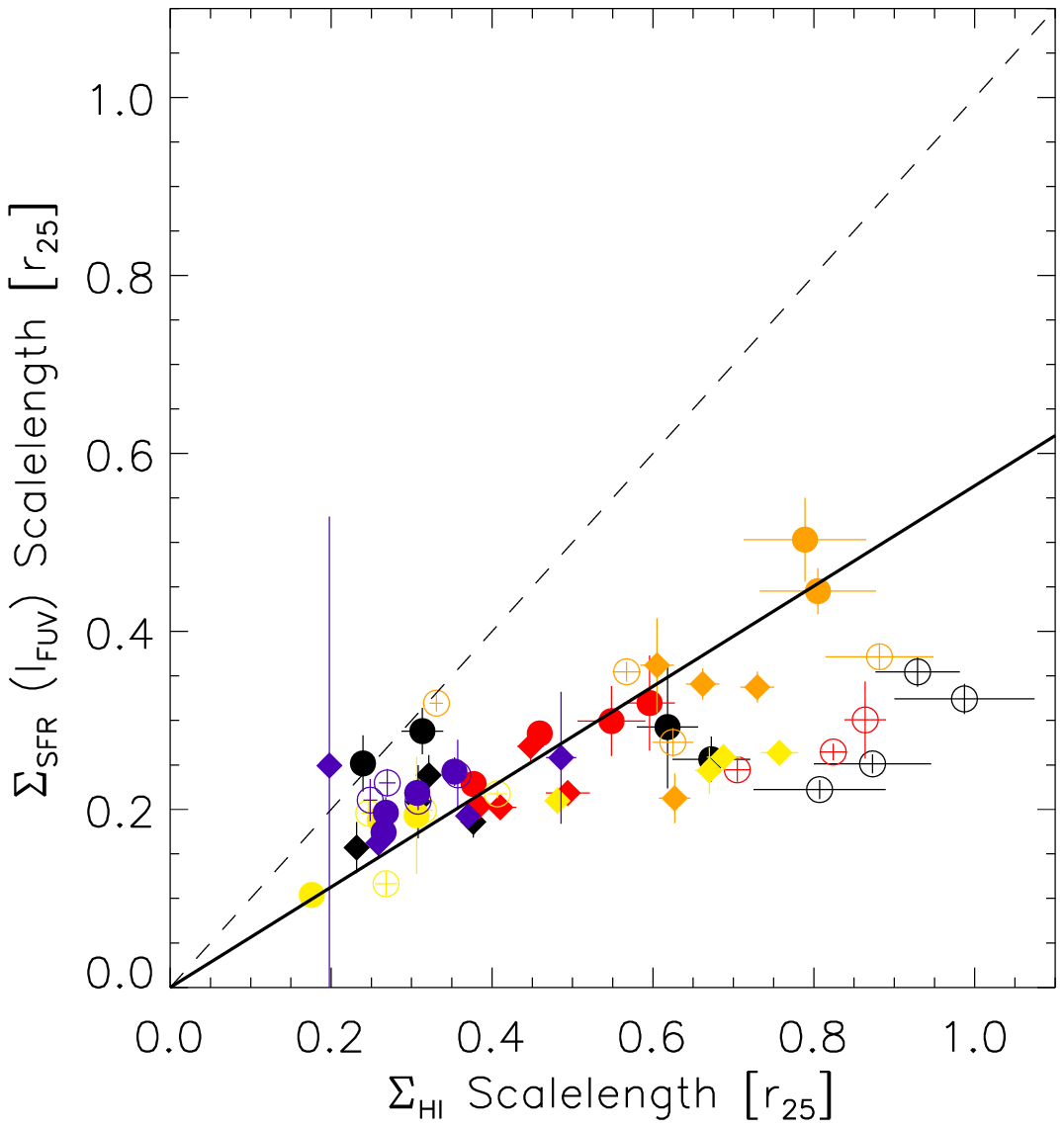}
\caption{Scale lengths for the exponential decline of FUV ($y$-axis)
  and {\sc Hi} ($x$-axis) in outer galaxy disks. Each galaxy contributes
  four independent measurements from different azimuthal sectors; data
  points that belong to the same galaxy share the same color and
  symbol. Black data points represent dwarf galaxies. The dashed line
  indicates a slope of unity. Virtually all data lie underneath this line,
  indicating that the FUV profiles generally decline more rapidly than the
  {\sc Hi} profiles. The solid line, which shows the mean ratio of the two
  scale lengths, has a slope of $\sim 0.5$, i.e., the outer disk {\sc Hi} scale
  length is twice as large as the FUV scale length on average.}
\label{fig3}
\end{figure*}

Both \hi\ and FUV emission decline as a function of radius across the outer disks.  Because
gas is the fuel for star formation, one would
expect the SFR to drop in a similar fashion compared to the \hi (if all other factors were equal).
To test this expectation, we derive radial profiles of each quantity and
characterize the decline at large radii using an exponential scale
length (i.e., the length over which the emission declines by a factor
$e$).

 Figure \ref{fig2} shows the radial profiles of \hi\ (black) and FUV
 (gray) together, along with the exponential fits, for our 22
 sample galaxies. The profiles are azimuthal averages over beam-wide
 ($15\arcsec$), concentric elliptical annuli constructed using
 inclination and position angles from Table \ref{table-general}. The
 error bar on each profile point indicates the RMS uncertainty in the mean in that
 ring. For easy comparison of profiles in different galaxies,
 we use common $\Sigma_{\rm HI}$  and $\Sigma_{\rm SFR}$ axes.
 The vertical dotted line indicates the innermost data point included in the
 fit (which is the profile point closest to $r_{25}$).

Horizontal lines indicate the sensitivity limits:
$0.09\,{\rm M}_{\sun}{\rm pc}^{-2}$ for $\Sigma_{\rm HI}$ (black), a value seldom
reached within $2\times r_{25}$, and the sensitivity ($2\,\sigma$) of
each FUV map, converted to units of $\Sigma_{\rm SFR}$ (gray). To
estimate these sensitivity limits we use the \hi\ sensitivity from
\S\,\ref{data-hi} and the FUV sensitivity from each map
and assume (a conservative number of) 20 independent
measurements contributing to each annulus to
reflect the gain in sensitivity by azimuthal averaging.

Solid lines show our exponential fits to $\Sigma_{\rm SFR}$ (gray)
and $\Sigma_{\rm HI}$ (black). We carry out the fits between $\sim1$
(the vertical dotted line) and $2\times r_{25}$ and consider only
profile points above our sensitivity cuts. We reiterate that for
NGC~5457 and NGC~5236 the field-of-view of our \hi\ maps is
too small to probe out to $2\times r_{25}$. We thus blank their radial
profiles outside their respective limiting radius (compare \S\,\ref{data-hi}).
We note that NGC\,3627 has too few data points above the \hi\
sensitivity cut to fit the profile beyond $r_{25}$. We also note that 
in particular NGC\,5236 (M83) is known to have extended FUV features in the
outer disk \citep[e.g.,][]{thilker05}. These features have a very low 
filling fraction for the outer disk annuli and are thus
not easily visible in our radial profiles. This galaxy is subject of a separate paper where
significantly deeper and more extended data are used to assess the relationship between \hi\ and
FUV emission out to many optical radii \citep{bigiel10}.

To fit the profiles we use an ordinary least-squares
(OLS) approach and estimate the uncertainties in the resulting fit in
several ways: by adding noise and re-fitting, bootstrapping
(resampling with repeats), varying the radii used to define the fit,
and considering only a subset of azimuthal angles (rather than the
whole ring).

In fact, azimuthal variations within a galaxy appear to dominate the
uncertainty in the fit, which might be expected from the common
appearance of streamers, tidal features, and outer arms in both the
\hi\ and SFR maps. To separate out this effect, we derive four
independent scale lengths for each galaxy, one for each of four
sectors with a 90 degree opening angle. The remaining factors
(bootstrapping, radius definition, noise) contribute to an uncertainty
in the scale lengths for each sector. The typical scatter among the
four sector scale lengths is $\sim0.1\times r_{25}$, with
significantly less scatter in the FUV profiles than the
\hi\ (indicating a higher degree of azimuthal symmetry in the FUV
emission than in the \hi ).

Figure \ref{fig3} shows \hi\ scale length as a function of FUV scale
length. Each galaxy contributes four points, one from each
$90\arcdeg$-wide sector, with points for a galaxy sharing the same
color and symbol. Black
points represent dwarf galaxies, otherwise color is arbitrary. Four
galaxies (DDO~154, NGC~2841, NGC~3198, NGC~3621) and 9
individual sectors are omitted from this
plot because their \hi\ profiles are too flat to be fit robustly over
the range we consider. Their profiles do appear to decline roughly
exponentially, but the derived scale length is $> 1 \times r_{25}$,
yielding too small a dynamic range for a robust fit . 
NGC~3627 is also omitted as there are too few data points above the
sensitivity limit to fit the \hi\ profile (see above).

Figure \ref{fig3} demonstrates that in almost every outer disk quadrant we
study, the surface density of star formation traced by FUV emission
drops more quickly than the surface density of gas (for comparison,
the dashed line shows equality). An exception is
one sector in NGC~925 (filled purple diamond) which shows a slightly
larger FUV scale length (though with large associated uncertainties). Even though generally
there is a steady decline in both
\hi\ and FUV with radius (Figure \ref{fig2}) and a clear correlation
between the scale lengths of the two declines, these declines are
not identical. Instead star formation drops with roughly half the scale
length of the \hi . This is illustrated by the solid line, which shows the
mean ratio of FUV scale length and \hi\ scale length of $0.5$ (the
$1\sigma$ scatter about the mean is $\sim 0.2$). Therefore, and despite
  the remarkable extent of star formation revealed by GALEX, it is
  worth bearing in mind that this widespread UV emission is still
  relatively centrally concentrated compared to the \hi .

\subsection{Local Correlation}
\label{rankcorr}

\begin{figure*}
\plottwo{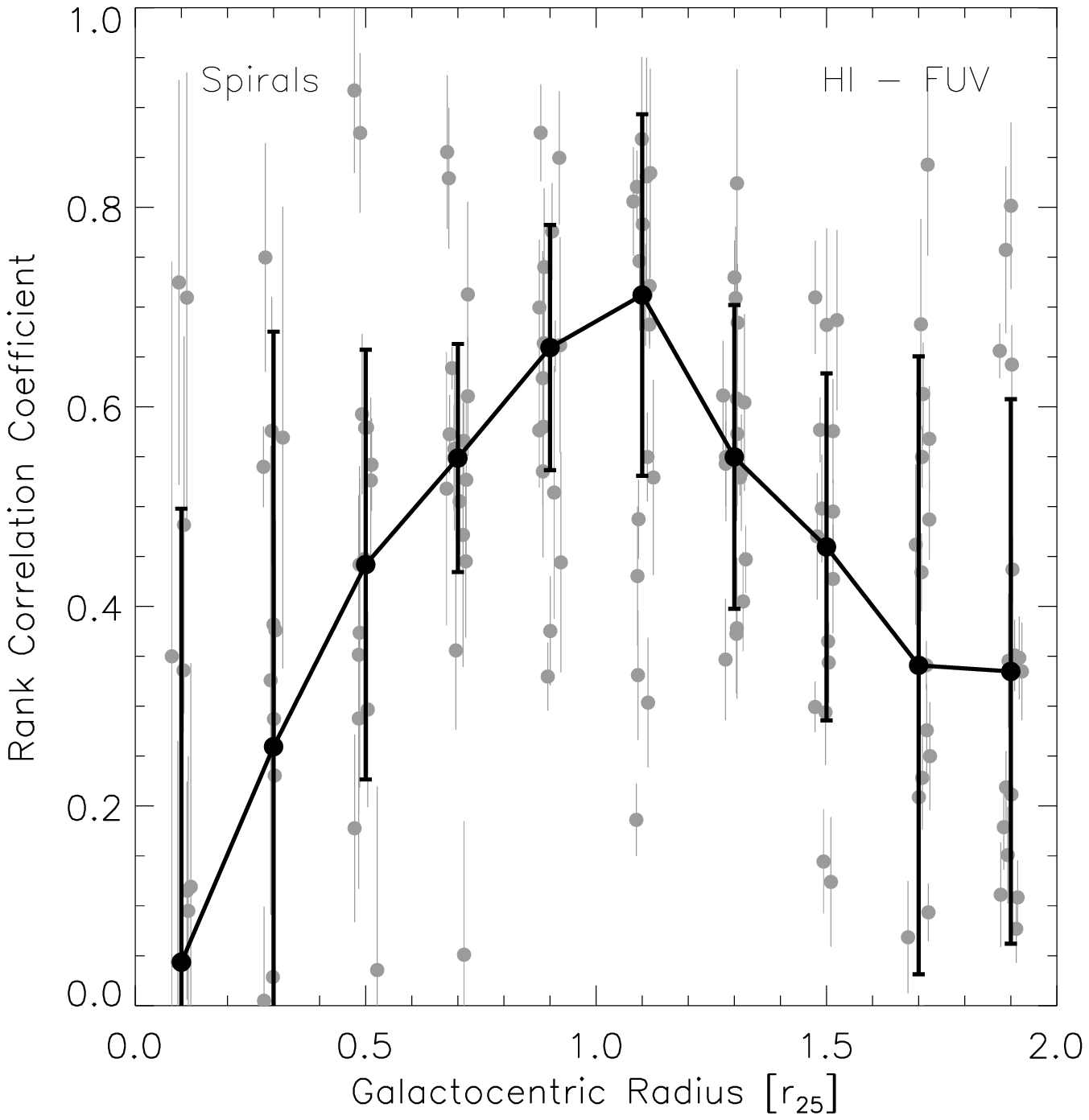}{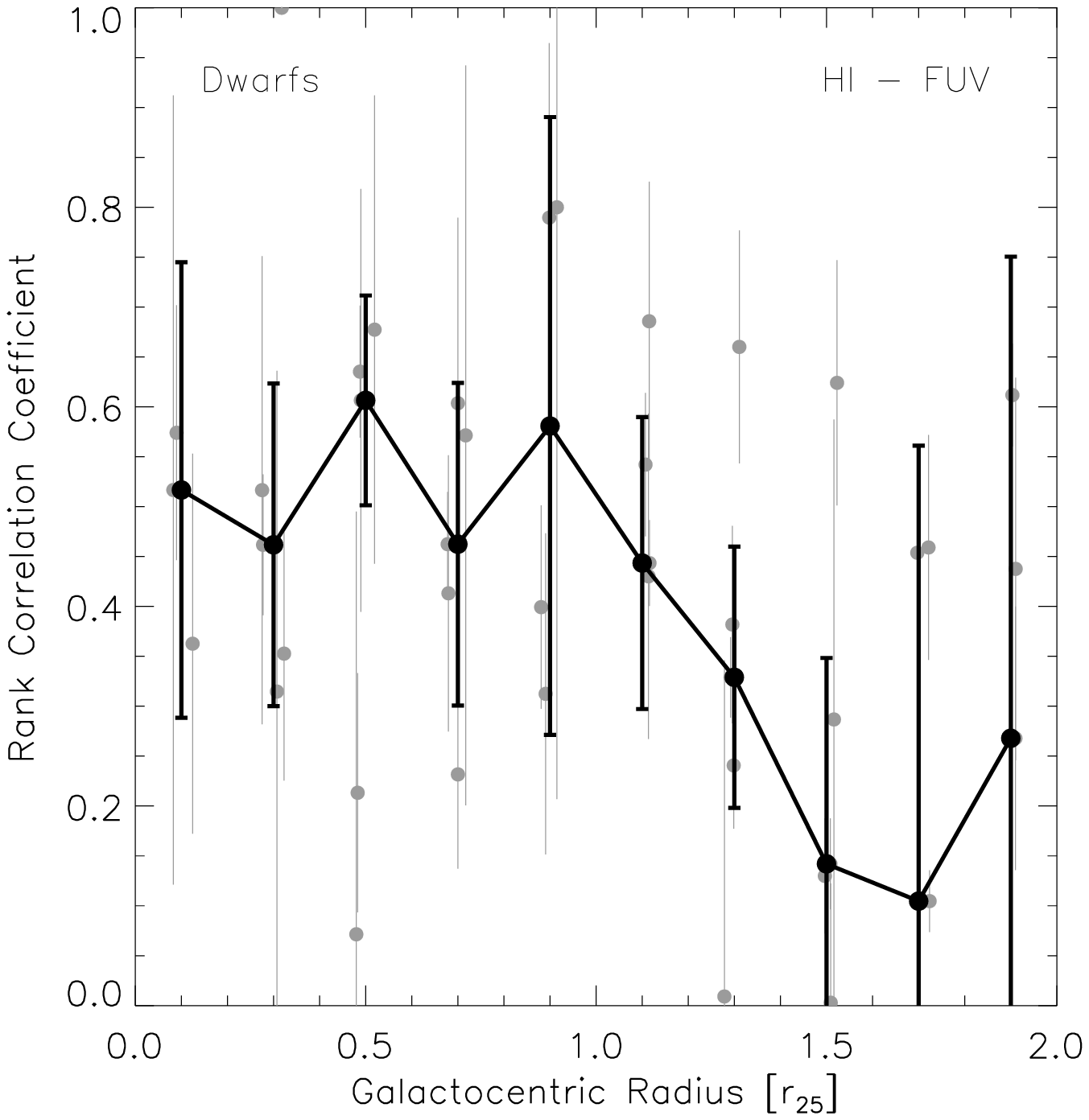}
\caption{Spearman rank correlation coefficient, $r$, relating {\sc Hi} and
  FUV as a function of galactocentric radius in the spiral (left) and
  dwarf (right) galaxies. For each galaxy, we measure $r$ in 10 radial
  bins spaced evenly between the galaxy center and
  $2\times$r$_{25}$. Each of these measurements, $r$ for one radial
  bin in one galaxy, appears as a gray point. Vertical error bars show Monte Carlo
  uncertainty estimates. The black curve in each panel indicates the
  median $r$ over all galaxies in each radial bin. Black error bars
  show the $1\sigma$ RMS scatter among galaxies in that bin. 
  FUV and {\sc Hi} emission are significantly
  correlated, with $r \gtrsim 0.5$, around the optical radius in
  the spirals and throughout the optical disks of the dwarf galaxies. In the inner parts of
  the spirals extinction and the (unaccounted for) presence of H$_2$ weaken the
  correlation. Outside the optical radius, $r$ decreases with
  increasing radius in both dwarfs and spirals.}
\label{fig4}
\end{figure*}

Comparing radial profiles tells us about the bulk behavior of \hi\ and
FUV in outer disks, but does not reveal information about any local
relationship. To assess the detailed relationship between FUV and \hi
, we now turn our attention to how the two quantities correlate on a
line-of-sight by line-of-sight basis at a given galactocentric
radius. Our approach is to divide each galaxy into 10 equally wide
radial bins between the galaxy center and $2\times r_{25}$ and then compute
the Spearman rank correlation coefficient between FUV and \hi\ in each
bin. The rank correlation coefficient \citep{press92}, $r$,
is a non-parameteric measure of the strength of any one-to-one
relationship between two quantities. Possible values range from $r=
-1$ to 1, with $r=1$ indicating a perfect correlation --- i.e., that
the brightest data point in \hi\ is associated with the brightest data point in FUV,
the second brightest in \hi\ with the second brightest in FUV, etc.
--- but giving no information on the functional form. On the
other hand $r=0$ indicates a lack of correlation (expected if the two
distributions are independent), while a perfect anti-correlation will
yield $r=-1$.

Figure \ref{fig4} shows $r$ as a function of normalized galactocentric
radius for the spiral (left) and the dwarf (right) galaxies separately. Each
gray data point represents $r$ measured for one galaxy in one $0.2\times
r_{25}$-wide radial bin (for better visibility we add a small amount of noise to the
$x$-position to distinguish individual measurements in the same bin
from one another). The vertical error bars show the scatter in $r$ derived from
randomly pairing \hi\ and FUV data for that bin; this process destroys
any correlation, yielding $r\approx0$ and the scatter observed repeating the
process 100 times gives us an estimate of the uncertainty (this is
most rigorously thought of as how confident we are that $r \ne
0$). The black line
condenses the gray points into a single trend by giving the median
correlation coefficient for all galaxies in each radial bin with error
bars indicating the $1\sigma$ RMS scatter among galaxies.

For spiral galaxies, $r$ rises from $\sim0.1$ near galaxy centers to
$\sim0.7$ at about $r_{25}$ and then drops again to $\sim0.3$ near
$2\times r_{25}$. To assess whether the low values of $r$ at very small
and very large radii still represent statistically significant
correlations, we extend the procedure that we used to estimate the
errors for the individual measurements. We randomly pair \hi\ and FUV
data (now for all spirals and dwarfs, respectively),
so that $r=0$ by construction, and bin these data and then
derive $\sigma$ from the scatter in $r$ about the known true value of
0. This exercise leads us to estimate $1\sigma \lesssim 0.04$ for the
median profiles in both the spiral and dwarf sample, implying that the
median $r$ is significantly different from 0 in the outer parts of
spirals but is consistent with 0 within the uncertainties in the inner
parts of spirals and the outer parts of dwarf galaxies.

Inside $\sim0.2\times r_{25}$, the distributions of \hi\ and FUV are
compatible with the hypothesis that $r=0$, i.e., that the two quantities
are unrelated. This regime is not the focus of this study, but the
result is easy to understand given that we neglect internal
extinction, which is significant in this regime, and molecular gas,
which dominates the gas budget in the ISM at these radii.

The effects of molecular gas and internal extinction should be
negligible in outer disks. The decrease in $r$ from its peak near the
optical radius towards larger galactocentric radii appears to
represent an intrinsic weakening of the correlation between star
formation and total gas in this regime.

Maps of both FUV and \hi\ in outer galaxies are often dominated by
tidal streamers, arms or other relatively narrow but extended
features. One way that the result in Figure \ref{fig4} might arise
even in the face of a one-to-one correlation between star formation and
gas is if such features exist but are systematically offset between
the \hi\ and FUV maps. To check this possibility, we repeat the
analysis in this section with a set of maps convolved to
$30\arcsec$ resolution. If a small systematic offset were driving the behavior at
large radii in Figure \ref{fig4}, we would expect $r$ to increase at
lower resolution. We do not find such an effect, suggesting
that systematic offsets in otherwise perfectly matched maps do not
drive the observed decline in $r$ with radius.

Dwarf galaxies (right panel, Figure \ref{fig4}) do not exhibit the
depression in $r$ at small radii seen in spirals, which can be explained because
extinction and the contribution of H$_2$ to the total gas budget are generally
small even in the inner disks of dwarf galaxies. It is not clear if the slightly lower $r$
at large radii found in dwarfs compared to spirals is real or a statistical artifact:
our sample of dwarfs is small (5) compared to our sample of spirals
(17) and the galaxies themselves are smaller, leading to fewer independent
line-of-sight measurements per galaxy. The outer disks of
dwarf galaxies do show, however, a decline in $r$ with increasing radius similar
to that seen in the spirals.

\subsection{The Star Formation Efficiency in Outer Disks}
\label{sfe}

We have seen that \hi\ and FUV emission exhibit different radial
behaviors (Section \ref{radprofs}) and that the local correlation between the
two declines from good ($r \approx 0.7$) near $r_{25}$ to poor
($r \approx 0.3$) near $2 \times r_{25}$. The first result means that the
rate of star formation {\em per unit gas} changes systematically
across outer galaxy disks while the second implies that quantities
other than the local gas reservoir may be important to set this
quantity. Here we directly investigate the variation of the rate of
star formation per unit gas with \hi\ column and radius in outer
galaxy disks. Following convention for extragalactic studies, we refer
to the star formation rate per unit gas (here FUV-per-\hi ) as the
star formation efficiency (SFE). This is the inverse of the gas
depletion time ($\tau_{\rm dep}^{-1}$), the time required for present
day star formation to consume the available gas reservoir; it is
sometimes quoted as a true (dimensionless) efficiency by normalizing
to $10^8$~yr (i.e., quoting the fraction of gas consumed every
$10^8$~yr). The three formulations are all equivalent within a
constant. All SFEs (depletion times) quoted in this paper include a factor of
$1.36$ to account for heavy elements. 

In the remainder of this section, we analyze a data set of matched
\hi\ and FUV intensities, each measured over an individual line of
sight at $15\arcsec$ resolution
and together covering the area between
$1$ and $2\times r_{25}$ for our whole sample. We extract these data from
the maps following the approach used by \citet{bigiel08} to study star
formation inside the optical disk. Briefly, we measure the intensity
from the maps (convolved to $15\arcsec$ resolution) at
non-overlapping sampling points (i.e., separated by approximately a beam width)
and spaced evenly to cover the area between $r_{25}$ and $2\times r_{25}$
in each galaxy. Each data/sampling point is assigned a weight equal to the inverse
of the total number of data points for the galaxy from which it is drawn. We
apply these weights when combining data in order to give equal weight
to each galaxy; this avoids a few large galaxies dominating all of our
plots. Other details are as in \citet{bigiel08}.

\subsubsection{Distribution of \hi\ Columns}
\label{dist-hi}

\begin{figure*}
\epsscale{1.1}
\plotone{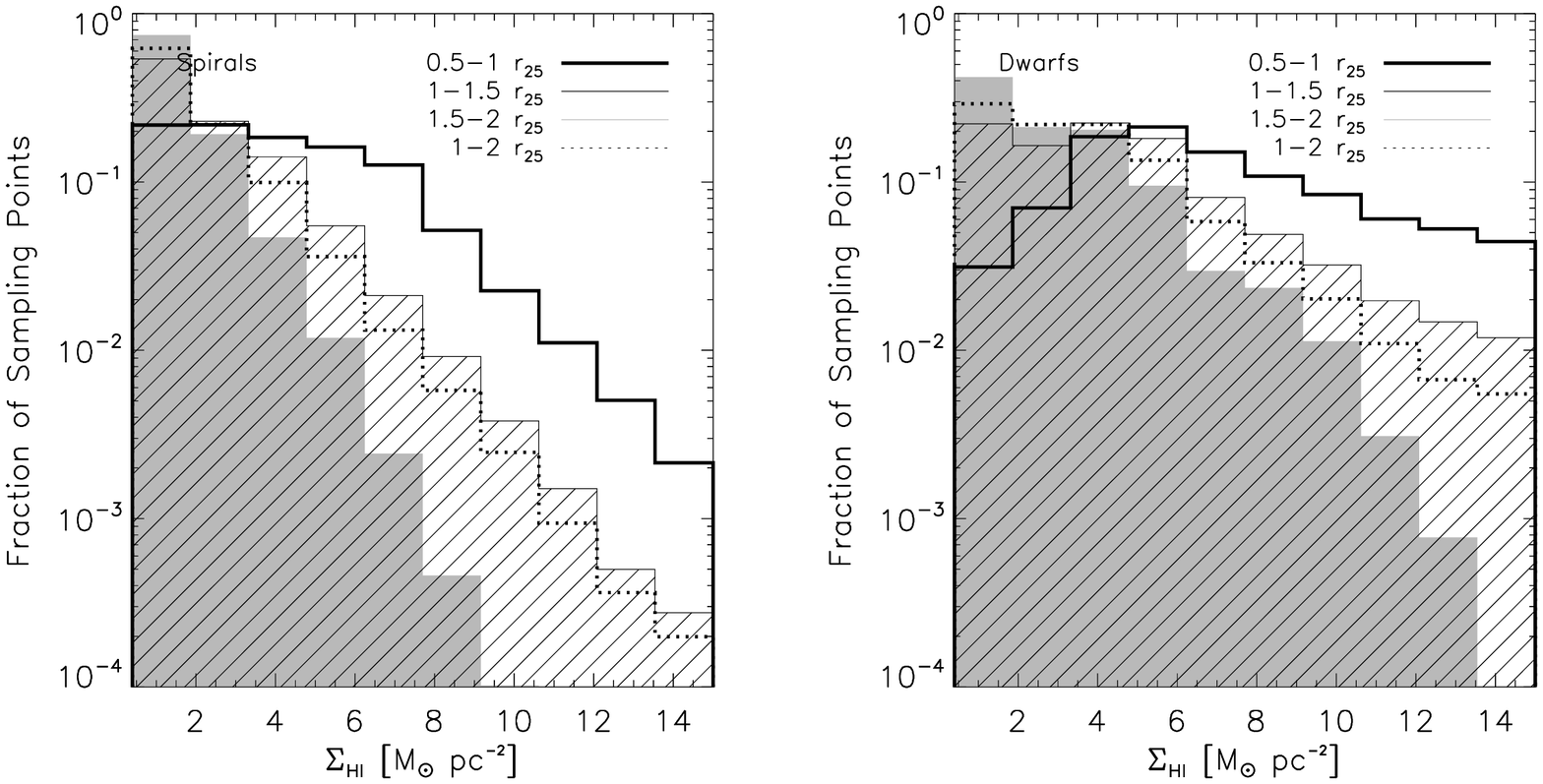}
\caption{The 4 histograms in each panel (left: spirals; right:
  dwarfs) show the normalized {\sc Hi} histograms between
  $1$ and $2\times r_{25}$ (black dotted line) and for 3 other
  radial regimes:  $0.5-1\times r_{25}$ (thick black line),
  $1-1.5\times r_{25}$ (black hashed) and $1.5-2\times r_{25}$ (filled
  gray). Every galaxy
  is assigned equal weight. In particular for the spirals, there is an approximately exponential
  decline in frequency with increasing $\Sigma_{\rm HI}$, with dwarfs
  showing a much shallower decline and more high-column data. 
  Each dwarf histogram is
  systematically shallower than the respective one for the spirals and
  with increasing radius, the fraction of sampling points with
  low {\sc Hi} columns increases strongly at the expense of those with
  high {\sc Hi} columns.}
\label{fig5}
\end{figure*}

Before we examine how the SFE depends on \hi\ column density, it will
be useful to see the actual distribution of \hi\ columns in our
data. We show this in Figure \ref{fig5} via
normalized $\Sigma_{\rm HI}$ histograms for the spiral (left panel) and the
dwarf (right panel) samples. The dotted histograms show the relative
distribution of \hi\ columns across the entire outer disks (between $1-2\times r_{25}$),
where we divided the \hi\ surface density range from $0.4 - 15\,{\rm M}_{\odot}~{\rm pc}^{-2}$
into 10 equally wide bins (there are very few data points at higher
columns and the lower limit corresponds to the sensitivity of the
\hi\ maps). In constructing the histograms, we give equal weight to
each galaxy rather than each data point. We test the robustness of the
histograms by repeatedly adding noise to the original data and
rebinning. Noise and false positives from regions below the
sensitivity cut only scatter the results by a few percent.

Unsurprisingly, the dotted histograms in Figure \ref{fig5} show that low \hi\ surface
densities dominate outer galaxy disks. The 50$^{th}$ percentile
surface density is $\sim1.6\,{\rm M}_{\odot}~{\rm pc}^{-2}$ for spirals and
$\sim2.3\,{\rm M}_{\odot}~{\rm pc}^{-2}$ for dwarf galaxies. The
spiral histogram appears approximately exponential, with an
$e$-folding every $\sim1.6\,{\rm M}_{\odot}~{\rm pc}^{-2}$.
The dwarf histogram, which is based on
fewer data, shows a shallower decline with a suggestion of a
flattening at low columns. Broadly, higher column densities are more
common in the outer parts of dwarf galaxies than in spirals.

We will also be interested in how the distribution of $\Sigma_{\rm
  HI}$ varies with radius. The other histograms in Figure \ref{fig5} show this. We divide the
data into three radial bins ($0.5$--$1.0$~$r_{25}$,
$1.0$--$1.5$~$r_{25}$, and $1.5$--$2.0$~$r_{25}$) and then separately
plot the same kind of normalized histograms for each bin. 

The relationship between dwarfs and spirals does not change
dramatically with radius. Spirals (left) show a steeper distribution
of $\Sigma_{\rm HI}$ than dwarfs (right) at all radii, so that dwarfs
always have more high-column \hi\ than spirals. Both spirals and
dwarfs do show a significant evolution in $\Sigma_{\rm HI}$ with
radius: The fraction of low \hi\ surface densities increases strongly
with increasing galactocentric radius and at the lowest radii,
$\Sigma_{\rm HI}$ flattens. The distribution even turns over for
dwarfs, so that low \hi\ columns are not the most common values
(instead $\Sigma_{\rm HI} \sim 4 - 6~$M$_{\odot}~$pc$^{-2}$ is the
most common value in dwarfs between $0.5$ and $1.0$~$r_{25}$).

\subsubsection{SFE vs. \hi\ Column}
\label{sfe-hi}

\begin{figure*}
\epsscale{0.6}
\plotone{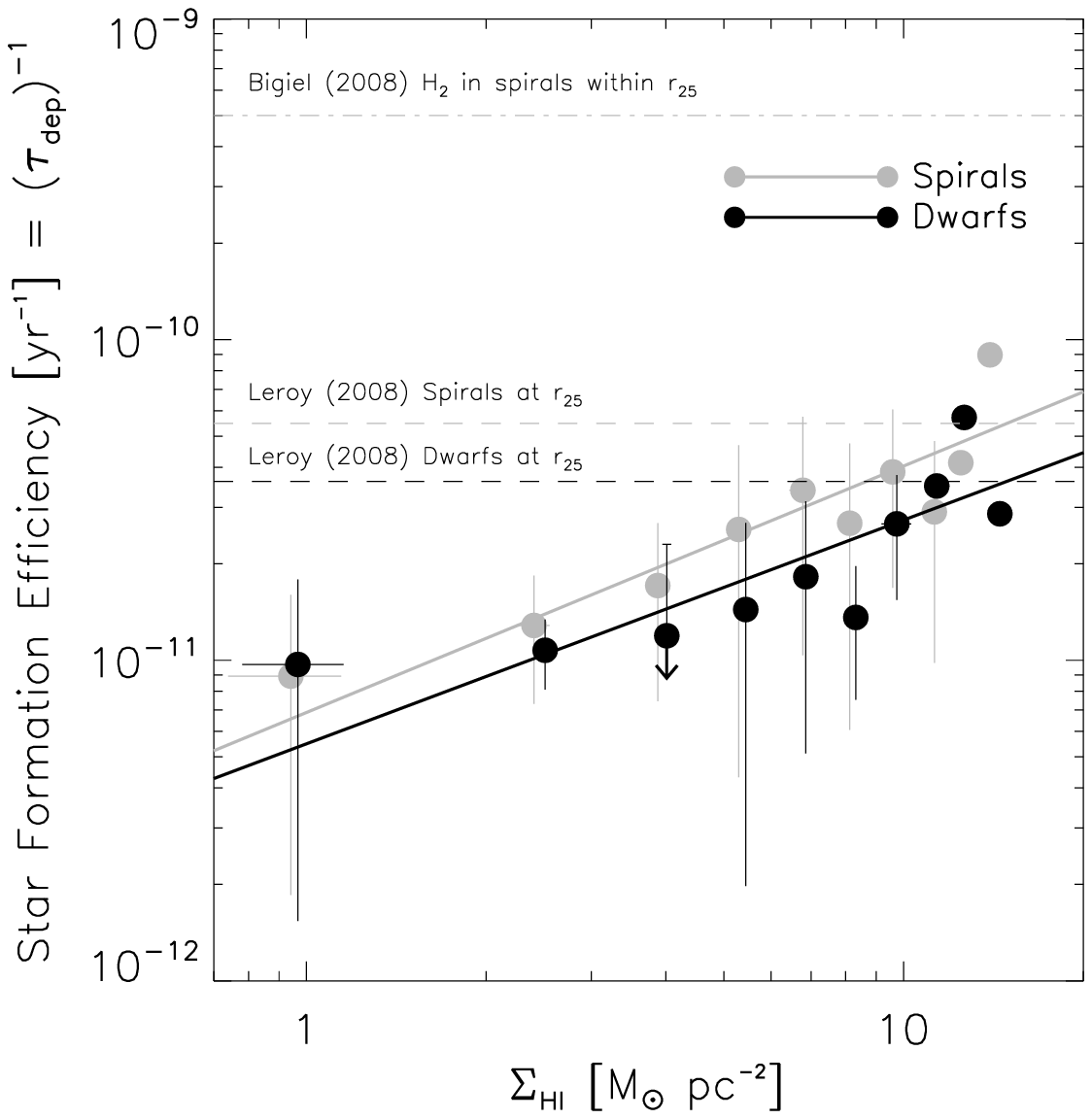}
\caption{SFE (i.e., SFR/gas or FUV/{\sc Hi}) binned by $\Sigma_{\rm HI}$ for both samples (spirals:
  gray; dwarfs: black). Individual filled circles represent the median SFE
  and $\Sigma_{\rm HI}$, error bars
  the $1\sigma$ RMS scatter in each bin (error bars are omitted on the
  highest two bins because only little data contribute). Every galaxy is
  assigned equal weight. The
  solid lines indicate power law fits to the two distributions (both
  cases have slope $\sim0.7$). Horizontal lines illustrate various
  comparison measurements: the SFE of molecular gas within the optical
  disks of spirals and of total (atomic plus molecular) gas at
  $r_{25}$ in dwarfs and spirals. The SFE is lower in outer disks than
  within the optical disk and rises with increasing {\sc Hi} column,
  showing a similar behavior in dwarfs and spirals.}
\label{fig6}
\end{figure*}

Figure \ref{fig6} shows the SFE as a function of $\Sigma_{\rm HI}$
for data between $1$ and $2\times r_{25}$. We
bin the data by $\Sigma_{\rm HI}$, assign equal weight to each galaxy
and plot the median SFE and $\Sigma_{\rm HI}$ in each bin. Error bars
indicate the $1\sigma$ RMS scatter among the data. Very few
data contribute to the highest column bins and as a result we cannot
robustly estimate the scatter there.

The measurements in Figure \ref{fig6} appear to all be statistically
significant. We check this by repeating the measurement after
substituting a pure noise map for the FUV map and by directly
propagating the statistical uncertainty in the two maps to derive the
uncertainty in the median for each bin. A pure noise measurement would
be well off the bottom of the plot and the formal uncertainty in each
bin is extremely small (as might be expected from the large number of
data).

The figure shows that aside from a slight offset the outer disks
of the dwarfs and the spirals exhibit a very similar SFE at a given
\hi\ surface density. In both cases, the SFE is a clear function of
$\Sigma_{\rm HI}$, increasing with rising \hi\ column density. Power
law fits to the points in Figure \ref{fig6} (solid lines) yield
indices of $0.7\pm0.1$ for both spirals and dwarfs. Although we saw
above that the distribution of $\Sigma_{\rm HI}$ does differ slightly
between the outer disks of spiral and dwarf galaxies, Figure
\ref{fig6} supports the idea that the outer disks of spirals and dwarf
galaxies represent similar regimes regarding the regulation and
timescales of star formation and that $\Sigma_{\rm HI}$ may be key to
this regulation in this regime.

Studying a subset of our spiral sample, \citet{leroy08} and
\citet{bigiel08} found a constant SFE for the H$_2$ that dominates the
gas reservoir in the inner part of spiral galaxies. The value measured
by \citet{bigiel08} appears as a dash-dotted horizontal line in Figure
\ref{fig6}. \citet{leroy08} further found the SFE of the {\em total} gas to
be a well-defined function of galactocentric radius in both spiral and
dwarf galaxies; we plot the SFE at $r_{25}$ predicted by their fit for
each subsample as dashed lines. Comparing our current data to these
lines we see that the SFE in outer disks is an order of magnitude or
more below that of the star-forming gas in the inner parts of galaxies, so that it will
take well in excess of a Hubble time to consume the gas reservoir {\em
  in situ}. The $r_{25}$ predictions from the \citet{leroy08} fits
coincide with the high end of our measured SFE, suggesting that the
steady decline they find may continue past $r_{25}$.

\subsubsection{Dependence on Radius}
\label{sfe-radius}

\citet{leroy08} and \citet{bigiel08} found that galactocentric radius
was actually a much better predictor of SFE than $\Sigma_{\rm gas}$ in
the inner parts of galaxies, likely because it tracks key
environmental quantities like the stellar potential well and
metallicity. To assess whether a similar radial gradient for the SFE holds in
outer disks, we separate our data into three radial bins ---
$0.5-1\times r_{25}$, $1-1.5\times r_{25}$ and $1.5-2\times r_{25}$
--- and repeat the above analysis (\S\,\ref{sfe-hi}) for each bin.

Figure \ref{fig7} shows the SFE as a function of $\Sigma_{\rm HI}$ for
each radial regime with dwarfs and spirals plotted separately. All
plot parameters are identical to Figure
\ref{fig6}. Again we plot no error bars if only a few galaxies
contribute to a bin. We also plot arrows instead of error bars where
the scatter exceeds the lower plot boundary. Reading Figure \ref{fig7}, it is important to bear in
mind that the SFE we plot between $0.5$ -- $1.0$~$r_{25}$ in spirals
is biased (in opposite directions) by our failure to account for
internal extinction or H$_2$ \citep[in these galaxies the ISM appears
  to be roughly equal parts \hi\ and H$_2$ at $\sim
  0.5~r_{25}$,][]{leroy08}; the points in Figure \ref{fig7} are
rigorously FUV-per-\hi\ rather than SFE.

In order to understand the upturn at low \hi\ columns, 
we traced the data for the lowest \hi-column bins back to
the original images and found that they mostly arise
from local \hi\ depressions (i.e., `\hi\ holes') in regions with diffuse, locally
smooth FUV emission. The FUV intensities involved are very low and inside a
galaxy disk -- the effect appears most pronounced for the
inner radius bin in Figure \ref{fig7}, i.e. within $r_{25}$ --
interpreting these intensities can be complicated; one
might expect some level of FUV emission from any galaxy due to an
intermediate age population, dust-scattered FUV light, or evolved
stars. Therefore, while Figure \ref{fig7} shows the real trend in our
data we emphasize the systematic uncertainty in the lowest bins,
particularly at low radii.

At higher $\Sigma_{\rm HI}$ ($\gtrsim 5$~M$_{\odot}$~pc$^{-2}$) all
radial regimes in both subsamples exhibit a positive correlation
between SFE and $\Sigma_{\rm HI}$ similar to what we saw in Figure
\ref{fig6} for the entire outer disks. The other general trend evident
for spirals is a decrease in SFE with increasing radius at fixed
$\Sigma_{\rm HI}$. For the dwarf sample the radial behavior is much
less clear. A decline in FUV per \hi\ with increasing radius is
clearly present at low $\Sigma_{\rm HI}$, but whether this represents
a real decline in SFE depends on one's interpretation of the low FUV
intensities.

\begin{figure*}
\epsscale{1.1}
\plotone{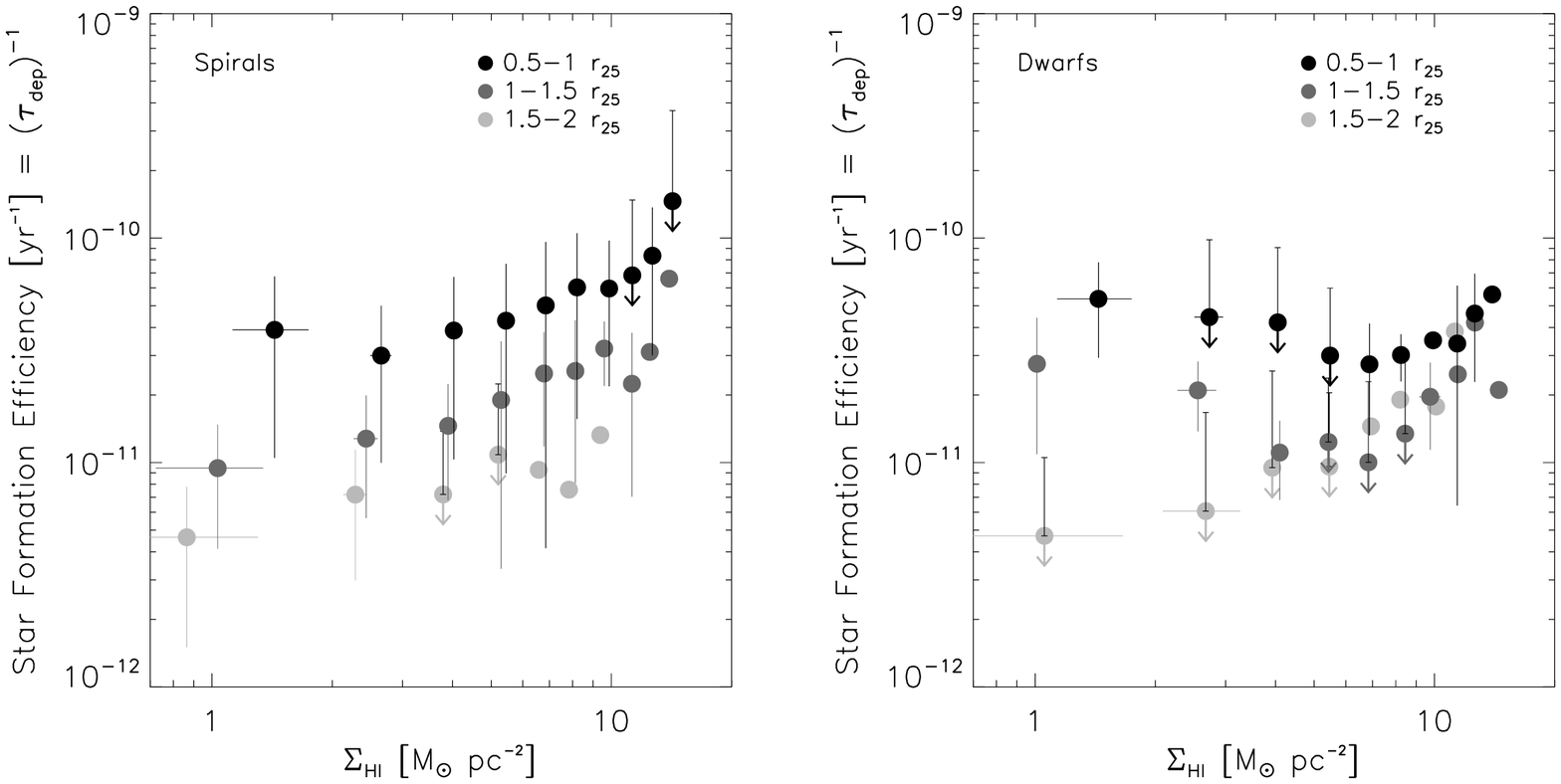}
\caption{SFE as a function of $\Sigma_{\rm HI}$ for spiral (left) and
  dwarf (right) galaxies. Methodology as Figure \ref{fig6} except that
  the data have now been divided into 3 radial bins with results for
  each bin plotted separately (black filled circles show radii
  $0.5-1\times r_{25}$, dark gray circles from $1-1.5\times r_{25}$
  and light gray circles from $1.5-2\times r_{25}$). We also
  plot arrows instead of error bars where
  the scatter exceeds the lower plot boundary. Generally, the SFE
  increases with $\Sigma_{\rm HI}$ and for a given $\Sigma_{\rm HI}$,
  the SFE decreases with increasing galactocentric radius.}
\label{fig7}
\end{figure*}

\subsection{Local Relations in Outer Disks}
\label{local}
\subsubsection{Star Formation and \hi}
\label{sflaw}

\begin{figure*}
\epsscale{0.9} \plotone{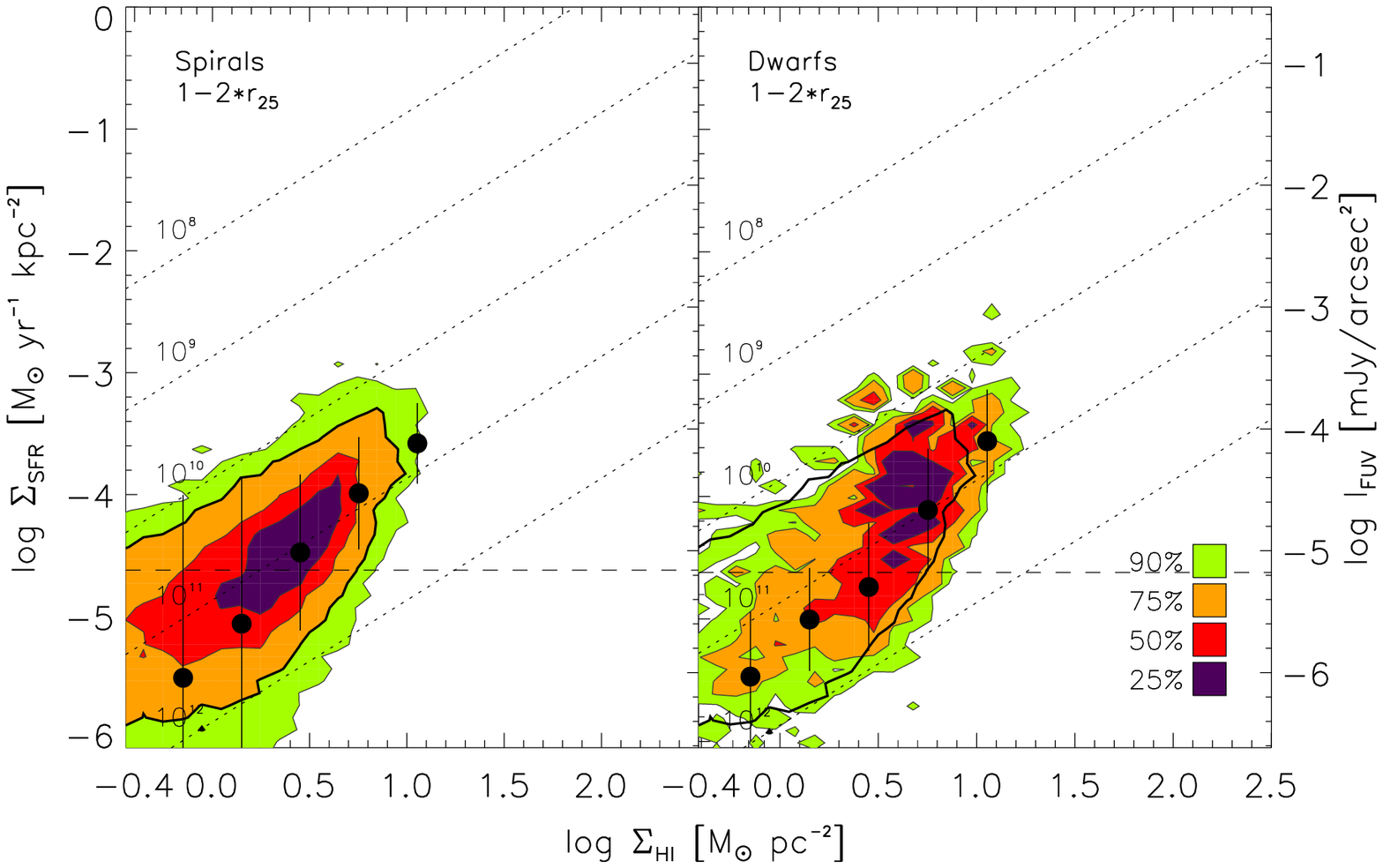}
\caption{Pixel-by-pixel distribution of FUV (right axis; left axis after
conversion to $\Sigma_{\rm SFR}$, Equation \ref{eq-sfr}) as a function of {\sc Hi} in
  the outer disks ($1-2\times r_{25}$) of spiral (left) and dwarf
  (right) galaxies. Contours show the density of data after combining
  all galaxies in each sample with equal weight given to each
  galaxy. Magenta, red, orange, and green areas show the densest
  $25\%$, $50\%$, $75\%$ and $90\%$ of the data, respectively. Dotted
  lines indicate constant {\sc Hi} depletion times of $10^{8}$ to
  $10^{12}$\,yr (taking into account heavy elements). A horizontal
  dashed line indicates the typical $3\sigma$ sensitivity of an
  individual FUV measurement. Black filled circles show our best estimate for
  the true relation between FUV and {\sc Hi} after accounting for finite
  sensitivity: they represent the median FUV after binning the data by
  $\Sigma_{\rm HI}$ and error bars are the lognormal scatter that
  yields the best match to the data after accounting for noise (see
  text). To allow easy comparison, we overplot the orange ($75\%$)
  contour for the spirals as a thick black contour in the dwarf
  (right) plot.}
\label{fig8}
\end{figure*}

We have so far binned and averaged our data to compare \hi\ and FUV in
a number of ways. It is also interesting to examine how individual
data are distributed in $\Sigma_{\rm SFR}$-$\Sigma_{\rm gas}$
(FUV-\hi) parameter space. Here we plot FUV as a function of
\hi\ directly for the data assembled in Section \ref{sfe}. This allows
ready comparison to work on the inner parts of galaxies by
\citet{bigiel08} and numerous literature measurements aimed at
constraining the star formation law \citep[e.g.,][]{kennicutt98}.

We have many thousands of data points, making a direct scatter plot
impractical. Therefore, we convert the data into two-dimensional
density distributions, again giving each galaxy equal weight. Figure
\ref{fig8} shows the resulting data density using contours to indicate
the area containing the densest 25\% (magenta), 50\% (red), 75\%
(orange), and 90\% (green) of the data. The data distributions in Figure \ref{fig8} and
the corresponding distributions inside $r_{25}$ are available as
supplemental online material.

Many of the conclusions from Sections \S\,\ref{dist-hi}
and \S\,\ref{sfe-hi} are again evident in Figure \ref{fig8}. Depletion
times are large (lines of constant \hi\ depletion time appear as
dotted diagonal lines in Figure \ref{fig8}) and change systematically
but relatively weakly with changing $\Sigma_{\rm HI}$. Dwarf galaxies
exhibit somewhat higher $\Sigma_{\rm HI}$ than spirals, leading to a
lack of low-column points in the right panel of Figure \ref{fig8}. At
a given \hi\ column density, the FUV one finds in spirals and dwarfs
is quite similar. This last conclusion can be clearly seen from the
right panel of Figure \ref{fig8}, where the orange contour from the
left panel appears as a thick black contour that closely matches the
distribution observed in dwarfs.

Sensitivity is a significant concern in this plot. The horizontal line
shows a typical $3\sigma$ sensitivity for our FUV maps. A large
fraction of our measurements lies below this line. This is problematic
for a log-log plot, where negatives are not reflected. To robustly
follow the general trend down to low $\Sigma_{\rm SFR}$, we
overplot median values for $\Sigma_{\rm SFR}$ in 5 equally spaced
$\Sigma_{\rm HI}$ bins as black circles. All data, including negatives,
contribute to the median, making it much more sensitive than each
individual point. Error bars on these points give our best estimate
for the intrinsic (log) scatter in $\Sigma_{\rm SFR}$ in each
\hi\ bin. We derive this estimate by comparing the observed data in
each bin to a series of mock data distributions. These are constructed
to have the observed median and appropriate Gaussian noise (measured
from the FUV maps) with varying degrees of lognormal scatter (from
$0.0$ to $2.0$~dex). We compare each mock distribution to the observed
data using a two-sided Kolmogorov-Smirnov (KS) test
\citep{press92}. The scatter that generates the distribution
most similar to the observed data is our best estimate for the true
intrinsic scatter.

For \hi\ columns $\gtrsim 3$~M$_{\odot}$~pc$^{-2}$, the scatter for
both spirals and dwarfs is a few tenths of a dex ($1\sigma$), so
that knowing the \hi\ column (and that one was in an outer galaxy
disk) could serve as an estimate for $\Sigma_{\rm SFR}$ with an accuracy of
a factor of 2-3. While this continues to be true at low columns in
dwarf galaxies, the scatter increases significantly below $\Sigma_{\rm
  HI} \sim 3$~M$_{\odot}$~pc$^{-2}$ for spirals (although the median
$\Sigma_{\rm SFR}$ for spiral and dwarf galaxies is still roughly the
same at these columns). If the observation is real and not an artifact
of poor statistics in the dwarf sample it may reflect a wider range of
environments in the outer disks of spiral galaxies.

One feature that particularly stands out in Figure \ref{fig8}, but was
not evident earlier, is an apparent maximum to the SFE in outer galaxy
disks. Almost all data have \hi\ depletion times longer than
$\sim10^{10}$\,yr (this line is a rough upper envelope to the green
contour in both panels). This implies that at their present SFR, even the most efficient outer disks in our
sample will not consume their available gas in less than 10~Gyr
(roughly a Hubble time) at their present SFR.

\subsubsection{SFE, Radius, and \hi}

\begin{figure*}
\plotone{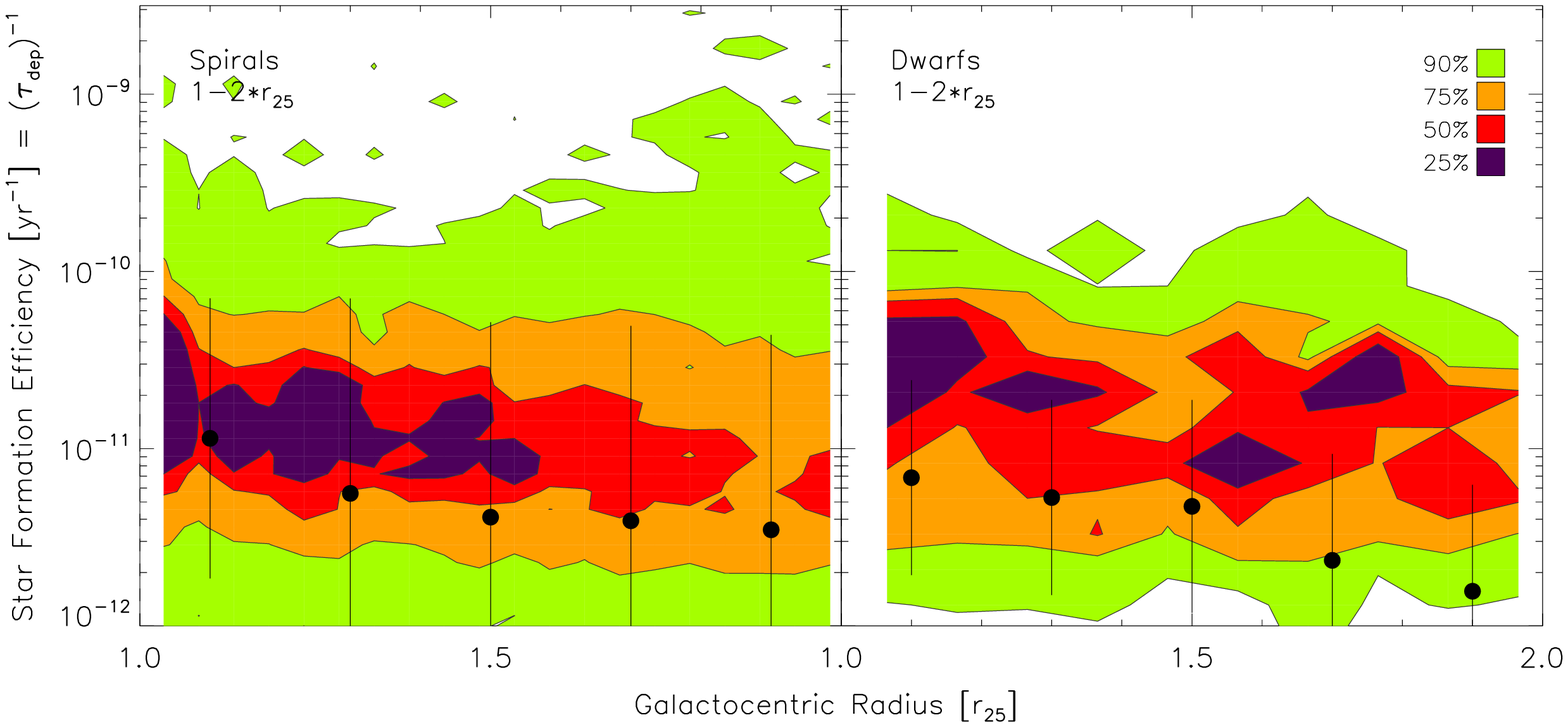}
\caption{FUV-to-{\sc Hi} ratio (SFE or $\tau_{\rm Dep}^{-1}$, $y$-axis) as a function of galactocentric
  radius in the outer disks of the spiral (left) and dwarf (right)
  galaxies. Contour levels, medians (black circles) and estimated
  intrinsic scatter (black error bars) are derived as for Figure
  \ref{fig8} (also compare text). As for the SFR-{\sc Hi} plots in Figure
  \ref{fig8}, both distributions are found to look almost identical,
  supporting the previous finding that the outer disk data from both
  galaxy samples lack clear distinguishing characteristics regarding
  their star forming properties. The SFE for both samples shows only mild
  variations across the outer disks.}
\label{fig9}
\end{figure*}

\begin{figure*}
\plotone{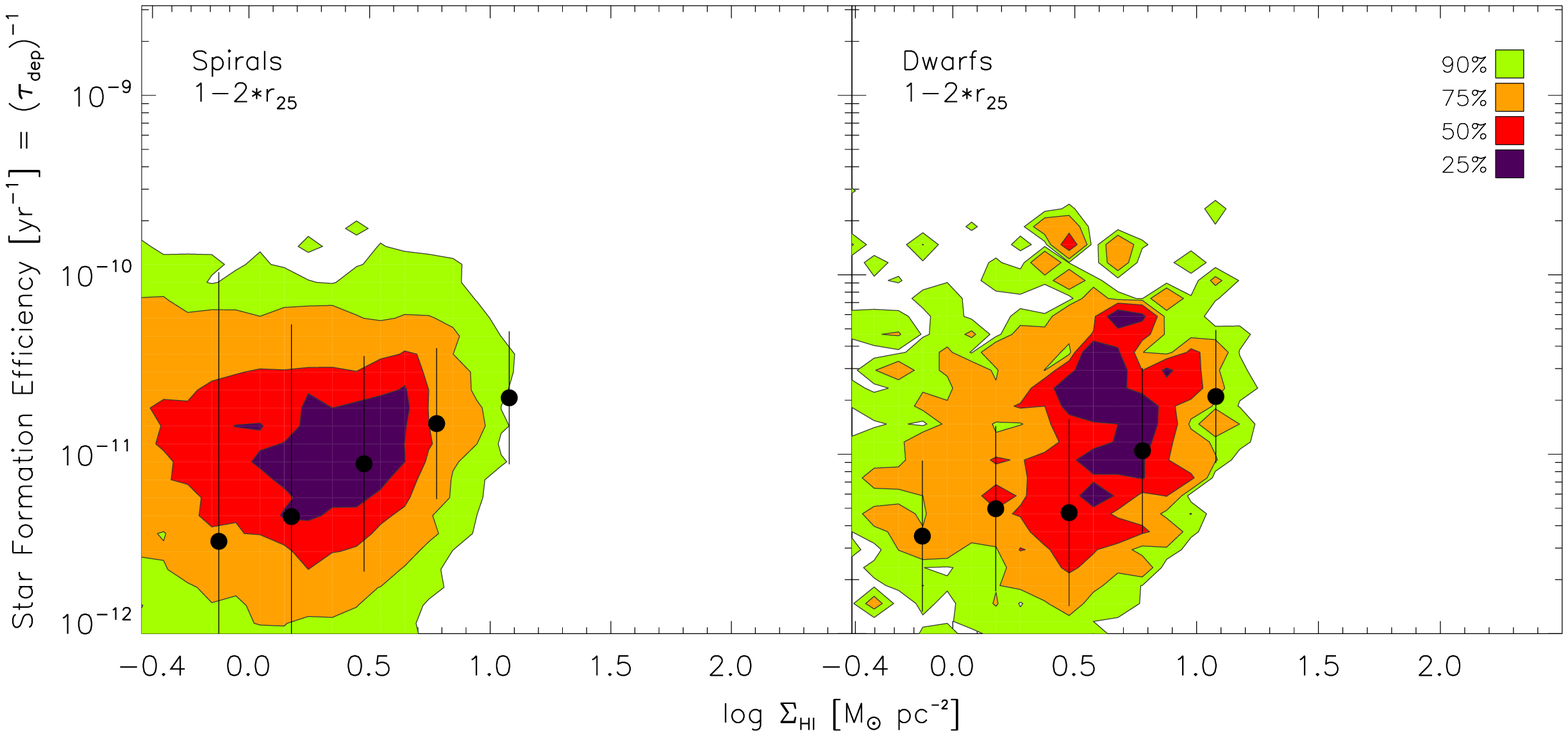}
\caption{The SFE ($\tau_{\rm Dep}^{-1}$) as a function of {\sc Hi} surface
  density $\Sigma_{\rm HI}$ for outer galaxy disks ($1-2\times
  r_{25}$) in spiral (left) and dwarf (right) galaxies. Contour
  levels, the methodology used to derive the median SFEs (black
  circles) and associated scatter, and other plot parameters are
  identical to the previous figures. We do not find an obvious
  discontinuity in the the SFE as a function of $\Sigma_{\rm HI}$ (as
  one might expect for a star formation threshold). Instead the
  SFE changes relatively smoothly as a function of $\Sigma_{\rm HI}$
  with significant scatter.}
\label{fig13}
\end{figure*}

In Figure \ref{fig9} we use the data to plot the SFE ($\tau_{\rm
  Dep}^{-1}$) as a function of radius. The
corresponding plot showing SFE as a function of \hi\ for individual
lines of sight appears as Figure \ref{fig13}. In both plots contour
levels and other details are identical to Figure \ref{fig8}. Black filled
circles and error bars show the median SFE and best-estimate intrinsic
scatter (using the same approach as above) in evenly spaced bins.

Again several earlier conclusions are more clearly visible in Figures
\ref{fig9} and \ref{fig13}: the SFE ($\tau_{\rm Dep}$) does appear to
vary systematically with both radius and \hi\ column, but this
variation is quite weak compared to the intrinsic scatter in the data
(especially for the radial variations in spirals). The difference
between the dwarf and spiral samples seems to be largely the
distribution of \hi\ columns: at a given column or radius the SFE
appears largely similar in the two samples and the range of SFEs found
for the two samples is similar ($\tau_{\rm Dep} \sim 10^{10}$ --
$10^{12}$~yr), with $10^{10}$~yr a rough lower limit to
$\tau_{\rm Dep}$.

These line-of-sight measurements also allow us to look for a
star-formation threshold, a set of local conditions below which star
formation is suppressed and above which it sets in. In a plot like in
Figures \ref{fig8} or \ref{fig13}, a true threshold as a function of
the ordinate should correspond to a rapid decline in the SFE as one
crosses the threshold. Neither Figure \ref{fig8} nor Figure
\ref{fig13} show clear evidence for such a behavior; the SFE varies
systematically but does not drop precipitously at any particular
\hi\ column density or radius. This does not necessarily rule out a
local threshold, but if one exists then the timescales and spatial
scales involved must be long enough and small enough to yield a smooth
trend of FUV/\hi\ as a function of \hi\ column and radius (albeit one
with large scatter).

\subsection{Toomre's $Q$ in Outer Disks}
\label{q}

Toomre's $Q$ parameter \citep{toomre64}, which measures the stability
of a thin axisymmetric disk (or ring, if the galactocentric radius is
kept as a free parameter), has been closely linked to the decline of
star formation in outer disks \citep[e.g.,][]{martin01}. Here we
compare $Q$ to the rate of star formation per unit gas in outer
galaxies to see if a universal behavior emerges. We calculate $Q$ via

\begin{equation} 
\label{eq-Q}
Q_{gas}=\frac{\sigma_{gas}\kappa}{\pi G \Sigma_{gas}}~,
\end{equation} 

\noindent with values $<1$ indicating instability (i.e., gas can
collapse and form stars) and values $>1$ indicating stability (i.e.,
no star formation). Here $\sigma_{gas}$ denotes the velocity
dispersion of the gas, $G$ is the gravitational constant and $\kappa$
is the epicyclic frequency. In the outer disks that we study, a flat
rotation curve is usually a good approximation \citep[see][]{deblok08}
and we can calculate $\kappa$ from $\kappa=1.41v_{flat}/r_{gal}$,
where $v_{\rm flat}$ is the rotation velocity at large radii. We adopt
$v_{\rm flat}$ from \citet{leroy08} where available and derive it in a
similar manner for galaxies that they did not study. We take
$\sigma_{gas} \approx 11\,{\rm km/s}$ everywhere, which is a
reasonable approximation for the \hi\ velocity dispersion (to within a
few km/s) over outer galaxy disks \citep[see][]{tamburro09}. Equation
\ref{eq-Q} assumes the mass budget to be dominated by gas
\citep[ignoring the stellar contribution considered by,
  e.g.,][]{jog84,rafikov01,yang07}; we consider this a reasonable
approximation outside $r_{25}$.

\begin{figure*}
\plotone{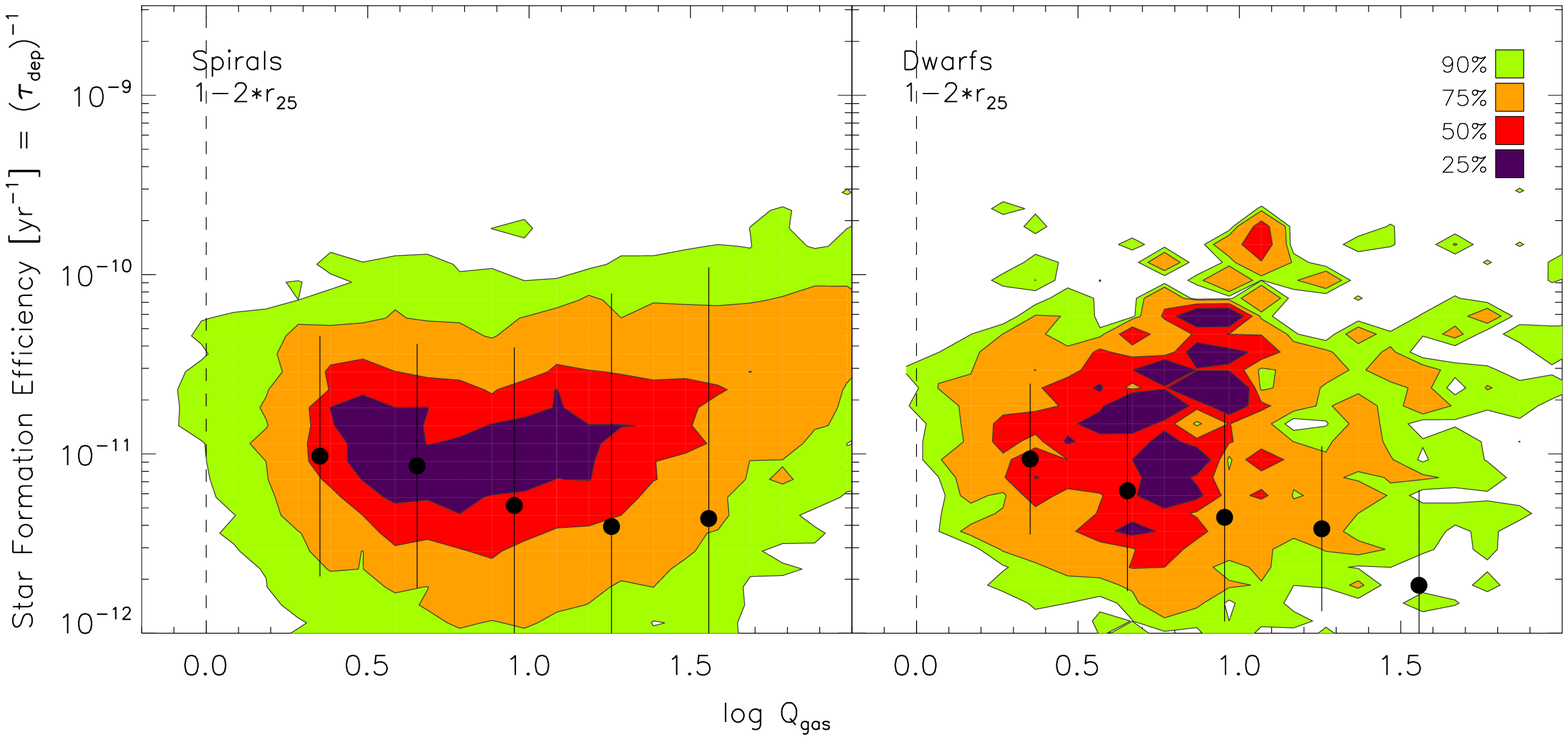}
\caption{FUV-to-{\sc Hi} ratio (SFE, $y$-axis) as a function of
  (log$_{10}$ of) Toomre's $Q$ parameter ($x$-axis) in the outer disks
  of spiral (left) and dwarf (right) galaxies. Details of the plot are
  as in the previous figures. $Q$ is calculated under the assumption
  of a flat rotation curve and $\sigma_{\rm HI} = 11$~km~s$^{-1}$
  \citep{tamburro09}. Almost all of the data are formally stable ($Q >
  1$, right of the dashed line). There is no clear trend in the
  FUV-to-\hi\ ratio (SFE) as a function of $Q$ at $\sim 750$~pc
  scales, as one might expect if $Q$ was the main criterion for a
  step-function type star formation threshold.}
\label{fig14}
\end{figure*}

Figure \ref{fig14} shows the FUV-to-\hi\ ratio as a function of
$Q$. Low values of $Q$ indicate gas that is more unstable to collapse
and might thus be expected to correspond to regions of more active
star formation per unit gas \citep[e.g.,][]{li05,yang07}. Such a trend
is not obvious from Figure \ref{fig14}, which shows that on the scales
we study most gas appears quite stable with little or no local
correlation between $Q$ and the FUV-to-\hi\ ratio.

Despite the lack of a strong local trend, there is an overall
correspondence: the depletion times that we derive are much larger
than those found inside $r_{25}$, while the typical values of $Q$ in
outer disks ($Q_{median}\approx30$) are, on the whole, much larger
than those found inside $r_{25}$. Inside r$_{25}$, \citet{leroy08}
found $Q_{median} \approx 4$ for the same methodology that we use here
and $Q_{median}\approx2$ considering a disk of gas and stars
\citep[also see][]{boissier03,yang07}.  Figure \ref{fig14} does not
offer clear evidence for Q as a `silver bullet' for star formation
thresholds, but outer \hi\ disks are clearly more Toomre stable than
gas inside $r_{25}$ \citep[for a discussion of the interplay between
  \hi\ phases, the gas velocity dispersion, and $Q$
  see][]{schaye04,deblok06}.
  
There are several subtleties to calculating $Q$: corrections are
sometimes applied for disk thickness (stabilizing) and the influence
of stars (destabilizing) and the appropriate \hi\ velocity dispersion is a
matter of some dispute. We use the median velocity dispersion measured
at $r_{25}$, but using a lower dispersion more appropriate to a cold
phase (e.g., $\sigma_{gas} = 6\,{\rm km/s}$ instead of 11\,km/s) would
not change our basic conclusions, as $Q$ would only decrease by a
factor of $\sim 2-4$ (even before accounting for the fraction of gas
in the cold phase, which must be relatively low given the observed
dispersions).

We do also note that there is a significant contribution of dark
matter to the local mass volume density in the outer disks. Assuming a flat
rotation curve, the local dark matter volume density $\rho_{dm}$ at
1.5\,$r_{25}$ is of the same order as the local \hi\ density, i.e.,
$\rho_{dm}\approx\rho_{HI}$. Deriving the effect of the dark matter on
the disk stability is non-trivial, but even if it were concentrated
into a thin disk along with the \hi\ it would not be enough mass to induce
formal instability.

\subsection{Comparison to Optical Disk Measurements}
\label{sflaw-r25}

We have looked at the relation between \hi\ and FUV in outer disks in
some detail. What remains now is to try to link these results to the
inner parts of galaxies. We do so in the following by comparing our results to
those of \citet{leroy08} and \citet{bigiel08}, who studied an
overlapping sample of spirals and dwarfs using many of the
techniques used here. It is absolutely essential to note that as we do so, we
abandon the clean mapping between physical quantities and observables
maintained throughout the first part of this paper: $\Sigma_{\rm gas}$ in
\citet{leroy08} and \citet{bigiel08} includes the contribution of
H$_2$ estimated from HERACLES CO maps \citep{leroy09} and
$\Sigma_{\rm SFR}$ represents a combination of FUV and IR (24$\mu$m)
emission, with the IR usually the dominant term (see those two studies
for details). With this caution firmly in place, we now compare the
scale length of star formation, the local relation between
$\Sigma_{\rm SFR}$ and $\Sigma_{\rm gas}$, and the SFE as a function
of radius across the two regimes.

\begin{figure*}
\epsscale{0.6}
\plotone{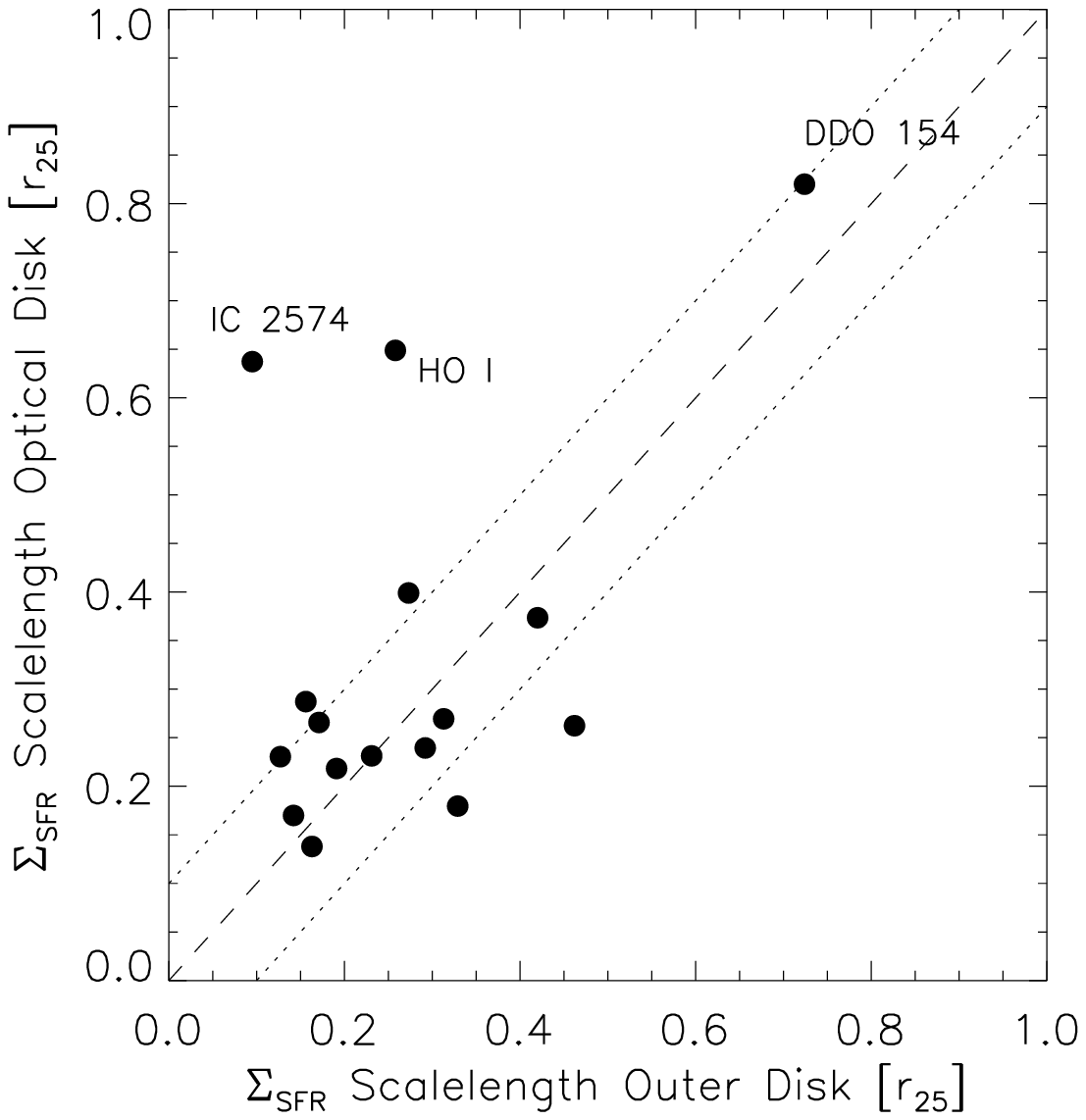}
\caption{Scale lengths of $\Sigma_{\rm SFR}$ (FUV + IR) measured
  within the optical disk by \citet{leroy08} as a function of our
  outer disk $\Sigma_{\rm SFR}$ (FUV) scalelengths for the overlap in
  samples. The dashed line indicates equality and the dotted lines
  show $\pm 0.1~r_{25}$. The two quantities agree reasonably well,
  with only the dwarf galaxies IC~2574 and Ho~I showing a sharp
  truncation in $\Sigma_{\rm HI}$ as well as in $\Sigma_{\rm SFR}$.}
\label{fig10}
\end{figure*}

Radial profiles of $\Sigma_{\rm SFR}$ appear to follow a steady
decline with approximately constant scale length from inner to
outer disks. We show this in Figure \ref{fig10}, where we plot scale
lengths fit by \citet{leroy08} to combined UV and IR profiles against
our outer disk measurements. There is significant scatter in the plot,
but with two notable exceptions the data appear to scatter around
equality with $\pm 0.1~r_{25}$ (typically $\sim 50\%$) scatter. The
two exceptions are the dwarf irregulars IC~2574 and Ho~I, which both
show a sharp downturn in their \hi\ and FUV profiles starting just inside
$r_{25}$ (see Figure \ref{fig2}).

By contrast, just glancing at Figure \ref{fig2} makes it clear that
the \hi\ does noticeably change its behavior between the inner and
outer disk. Our \hi\ profiles are usually fairly flat inside the
optical disk (sometimes with central depressions), while almost all of
the profiles show a radial decline outside $r_{25}$. We have already
seen that this decline is still shallow compared to the decline in
$\Sigma_{\rm SFR}$. \citet{leroy08} found that the CO and SFR scale
lengths were comparable inside $r_{25}$, so we expect that the decline
in \hi\ is also shallower than the decline in CO in the inner
disk. This suggests that the total gas (\hi\ + H$_2$) distribution may
follow a broken exponential rather than a single profile.

\begin{figure*}
\plotone{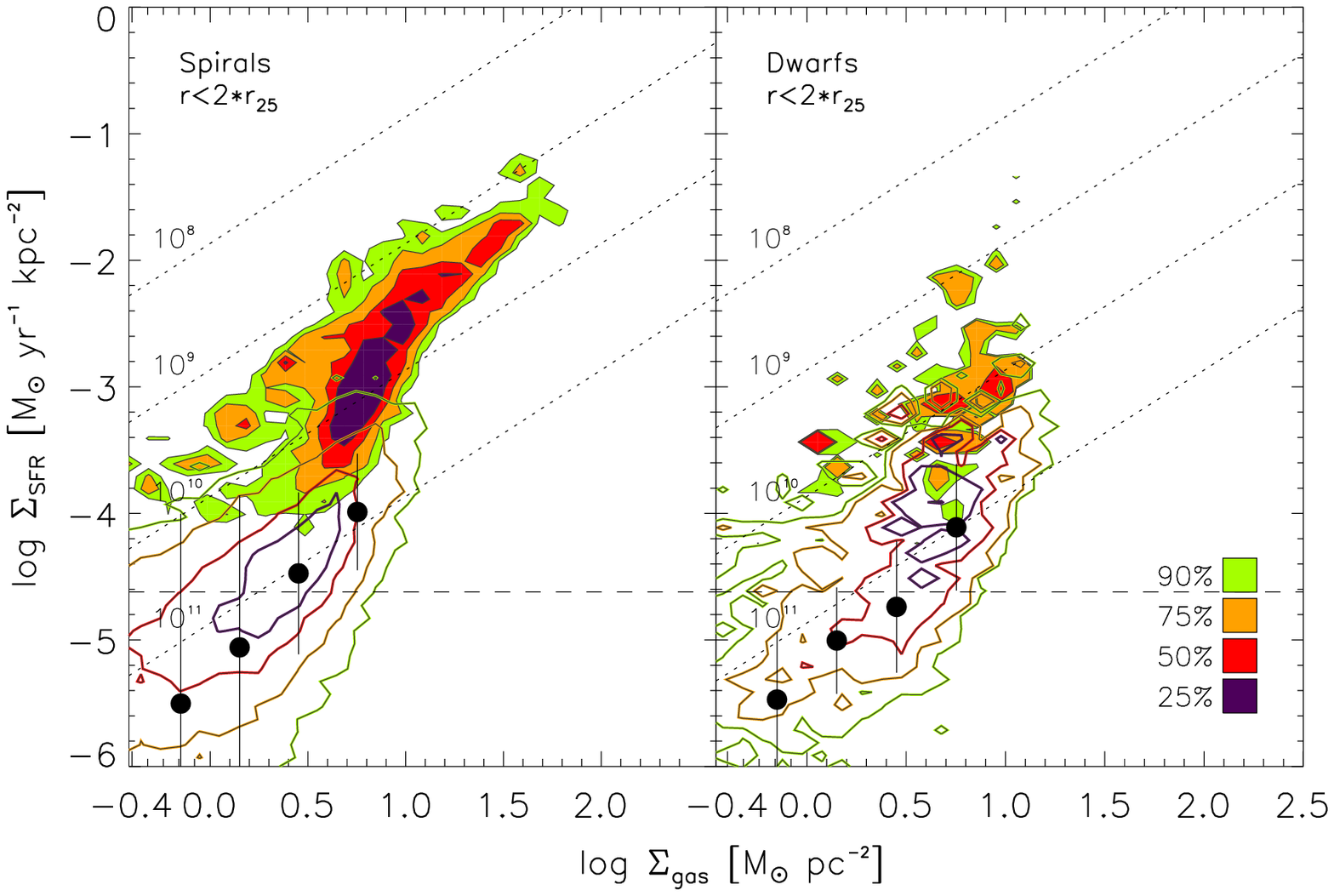}
\caption{Star formation and gas from inner to outer disks (left:
  spirals, right: dwarf galaxies). Filled contours show the
  pixel-by-pixel distribution of $\Sigma_{\rm SFR}$ as a function of
  $\Sigma_{\rm gas}$ inside $r_{25}$ (filled contours) for a subset of
  our sample \citep{bigiel08}. Empty contours show outer disk data
  (see Figure \ref{fig8}) with black filled circles indicating median and
  scatter. Plot parameters are as in Figure \ref{fig8}. In spirals,
  the steep relation between $\Sigma_{\rm SFR}$ and $\Sigma_{\rm
    gas}$ seen inside $r_{25}$ becomes more shallow at large radii,
  leading to relatively modest variations in the median SFE (depletion
  time) across the outer disks (though the scatter is large). For
  both spiral and dwarf galaxies, data from inside $r_{25}$ appear to
  more or less continuously extend into the outer disk. Note that the slight offset
  in $\Sigma_{\rm SFR}$ between the outer and inner disk data is due to different
  methodology in deriving SFRs in the two regimes (see text for details).
  These data are also available as supplemental online material.}
\label{fig11}
\end{figure*}

In Figure \ref{fig11} we compare individual line-of-sight measurements
between the optical and outer disks. Plot parameters largely match
Figure \ref{fig8}, though now the outer disk data appear as {\em unfilled}
contours (the median and estimated scatter are still black
points with error bars). {\em Filled} contours show data from within
the optical disk \citep[][their Figures 8 and 12 with slight changes
  to methodology: we assign equal weight to each galaxy, use a larger
  bin width, and define the contours via the fraction of data enclosed
  rather than absolute counts]{bigiel08}. Tables containing the
distribution and sampling data used to construct Figure \ref{fig11}
(optical and outer disk data for the spiral and the dwarf sample) are
available as electronic supplemental material.

The outer disks in Figure \ref{fig11} largely extend the distribution
found within the optical radius to lower SFR and gas surface
densities. There may be a small discontinuity between the two
distributions along the $y$-axis due to the inclusion of an IR-based
extinction correction inside the optical radius. What is striking is that the outer disk
data lie overwhelmingly in a different part of parameter space from
data inside the optical disk. In both samples, we see a smooth trend
extend to very long depletion times and low gas columns that are
almost totally absent inside the optical disks. Also in contrast to the
optical disks, a clear trend relating \hi\ and SFR emerges in both
subsamples. Inside $r_{25}$, particularly within $0.5\times r_{25}$, the
relationship between \hi\ and $\Sigma_{\rm SFR}$ is weak or
nonexistent; instead H$_2$ is clearly correlated with $\Sigma_{\rm
  SFR}$ \citep[][]{bigiel08}.

There appears to be a `forbidden region' in Figure
\ref{fig11} at relatively high $\Sigma_{\rm HI}$ ($\sim
3$--$10$~M$_\odot$~pc$^{-2}$ or log$\left(\Sigma_{\rm HI}\right)$ between
$\sim$\,0.5 and 1) and low $\Sigma_{\rm SFR}$.
This lack of high-\hi\ low-FUV gas combines
with the upper envelope of $\tau_{\rm Dep} \sim 10^{10}$~yr in the
outer disks and the
turn towards a fixed $\tau_{\rm Dep}$ in the H$_2$ dominated (inner) parts of
galaxies to create a combined distribution with an `S-like' shape
(though note that there is some gas at low columns with $\tau_{\rm
  Dep} \lesssim 10^{10}$~yr inside the optical disks of spirals). We
will interpret this combined distribution in Section \ref{discussion}.

The dwarf distributions lack the upper turn to the right in the
`S-shape:' we do not observe a significant amount of data at high
$\Sigma_{\rm gas}$, at least partially because we lack information on
the amount of \htwo\ in these systems. Instead we observe a soft upper
limit to the \hi\ surface density of $\sim 10\,M_{\sun}$~pc$^{-2}$ at our
resolution of $15\arcsec$
\citep[the `saturation' discussed by][]{bigiel08}. Figure
\ref{fig11} shows that this pile-up at the saturation extends into the
outer disks of dwarf galaxies, though the SFR for $\Sigma_{\rm HI}
\sim 10$~M$_\odot$~pc$^{-2}$ outside $r_{25}$ is lower than for the
same gas inside $r_{25}$, on average. Unless H$_2$ represents the
majority of the ISM by several times in dwarf galaxies, it seems safe
to conclude that as in spirals star formation is more efficient inside
the optical disks of dwarf galaxies than at large radii.

Finally, in Figure \ref{fig12}, we look at how the SFE varies as a
function of galactocentric radius from 0 to $2\times r_{25}$.  We plot
our data at $> r_{25}$ (right of the vertical dashed line) and
measurements by \citet{leroy08} inside $r_{25}$ (their Figure 1); we
also indicate their fits to the SFE as a function of radius for each
sample. As in the case of the SFR vs. gas plot (Figure \ref{fig11}),
some discontinuity reflecting the change in methodology is seen
around $r_{25}$ (we illustrate the effects of neglecting internal extinction
within $r_{25}$ by the black set of contours, which show an extension
of the red filled contour of the outer disk data towards
smaller radii). Our approach here is not appropriate to investigating
the detailed behavior around $r_{25}$, nor is this our goal; we wish
to broadly compare the inner and outer disk regimes.

Again, we see a distinct contrast between the inner and outer
disk. The SFEs overall are lower, the scatter in SFE appears larger,
and although the first few points in both samples are approximately
consistent with the decline measured by \citet{leroy08}, the overall
decline in SFE with radius across the range $1$--$2\times r_{25}$ is
markedly more shallow than between $0$ and $1\times r_{25}$.

\begin{figure*}
\plotone{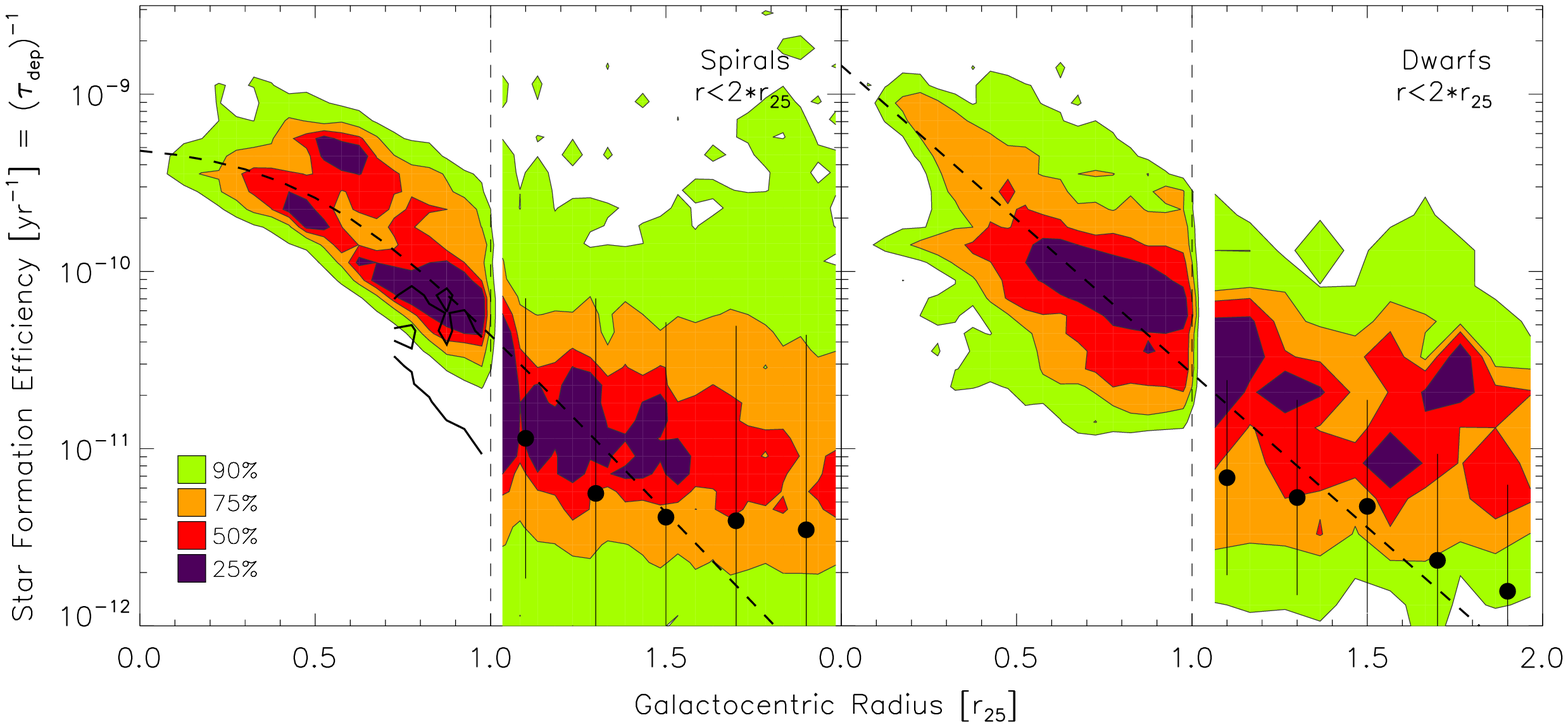}
\caption{SFE (i.e., SFR/gas or FUV/{\sc Hi}) as a function of galactocentric
  radius from 0 to $2\times r_{25}$ in spiral (left) and dwarf (right)
  galaxies. The vertical dashed line indicates $r_{25}$. Contours and
  black filled circles right of this line are as in Figure \ref{fig9}. Data
  inside $r_{25}$ are from \citet{leroy08} and are derived accounting
  for internal extinction and the presence of molecular gas (meaning
  that some discontinuity at $r_{25}$ may be expected). We illustrate the effects
  of neglecting these two factors by the black set of contours, which show an extension
  of the outer disk data (the red contour) towards smaller radii. The dashed
  lines show fits to the SFE inside $r_{25}$ from \citet{leroy08}. The
  outer disk data are roughly consistent with these trends inside
  $\sim 1.5~r_{25}$ but at large radii the decline in the SFE is more
  shallow than it is inside the optical disk.}
\label{fig12}
\end{figure*}

\section{Discussion}
\label{discussion}

\subsection{Broad Structure of Star Formation In Outer Disks}

Our most basic conclusion is that star formation in outer disks is
extremely inefficient compared to star formation inside the optical disks.
\hi\ depletion times for most of our data are
$\sim10^{11}$~yr and it is rare to find regions with local depletion
times $\lesssim 10^{10}$~yr. These values agree with previous
measurements near $r_{25}$ and observations of low surface brightness
galaxies \citep{wyder09}. Thus present day star formation requires
many Hubble times ($\sim 10^{10}$~yr) to devour the existing gas
reservoir. Another way to look at this is the low integrated SFR of
outer galaxy disks: on average, outer disks contribute only about
$10\%$ to the total SFR of a galaxy. The relative lack of importance of {\em in situ} star formation
means that the massive extended gas distributions observed in many
nearby galaxies can be long-lived and may be viewed as a potential source of
fuel for inner disk star formation. The short depletion times in inner
disks \citep[$\sim 2 \times 10^9$~yr,][]{bigiel08,leroy08} imply that
such a source is required \citep{shlosman89,blitz96,bauermeister09}, though the
presence and importance of radial gas flows are still debated
\citep[e.g.,][]{vollmer03,wong04,peek09}.

Our second simple conclusion is that both \hi\ column and FUV
intensity decline systematically at large radii. For \hi, this
represents a contrast with the inner disk, where the azimuthally
averaged \hi\ surface density tends to be flat or even increase with
increasing radius. As a result, outer disks are home to column
densities seldom found inside the optical radius; in the regime we
study in spirals, column densities $<3$~M$_\odot$~pc$^{-2}$ account
for the majority of data. For FUV, the decline that we observe outside
$r_{25}$ appears approximately consistent with the decline in SFR
inferred from combined IR and UV profiles inside $r_{25}$
\citep{leroy08}. The rate of decline in \hi\ and FUV intensity for a
given galaxy appears to be correlated but not identical;
parameterizing the decline with exponential scale lengths, we find the
FUV intensity to decline with a scale length approximately half that
of the \hi . With the SFR declining faster than the gas reservoir,
star formation becomes less efficient with increasing radius, though
this decline is actually milder than that observed inside $r_{25}$.

Despite the mismatch in scale lengths, we find a fairly good overall
relationship between \hi\ and FUV (measured via direct scatter plot or
rank correlation). This is surprising given the extremely poor
correlation between \hi\ and SFR (traced mainly by IR emission) found inside
star-forming disks \citep{wong02,kennicutt07,leroy08,bigiel08} and gives a
strong hint that different physics governs the formation of
star-forming clouds at large radii. This agrees qualitatively with
the results of \citet{hunter10}, who found that the integrated HI-richness
of dwarf galaxies appears related to outer disk star formation in the sense
that HI-poor dwarfs preferentially show suppressed star formation.

\subsection{Star Formation Thresholds}

Although we do find the SFE to depend on both gas column and radius,
we do not find clear evidence for a threshold in either quantity. This
agrees with numerous recent studies that have failed to detect such a
feature using UV data
\citep[e.g.,][]{boissier07,thilker05,hunter10,gildepaz05,gildepaz07b},
but is in contrast to the radial cutoffs found in \halpha\ emission
\citep[e.g.,][]{kennicutt89,martin01}. The main new contribution of
our work is to look for the signature of such a feature by plotting
the SFE of individual lines of sight as a function of \hi\ column
density. Doing so, we observe a continuous dependence of SFE on
\hi\ column with large scatter rather than an abrupt step at a given \hi\
column density.

What does the lack of a step function in FUV/\hi\ as a function of
\hi\ mean? Our angular resolution ($15\arcsec$) translates to $\sim
750$~pc for a typical spiral in our sample ($d \sim 10$~Mpc). A
reductionist view of our observations would imply that if there is a
critical column density that must be achieved on these scales in order
for star formation to occur, then this column density is not
maintained over the same timescales and area as the FUV emission from
the young stars that result. Our best estimate is that this is not due
to internal extinction or unaccounted H$_2$, though in both cases this
represents an assumption. It may be a timescale effect, with FUV
emission surviving its parent gas. FUV emission at low but significant
levels may be found from stellar populations as old as 100~Myr, during
which a population with a 10~km~s$^{-1}$ dispersion (a plausible
dispersion for young stars and roughly the observed dispersion in the
gas) might move $\sim 1$~kpc. This would allow FUV emission from one
resolution element in our line of sight analysis to diffuse to nearby
regions and the gas distribution to evolve significantly.

Another explanation is that star formation has only a
stochastic relation to kiloparsec scale conditions. In the inner parts
of galaxies, star forming clouds are $\sim 50$~pc in size, a factor of
10--20 times smaller than one of our resolution elements. A similar
contrast holds in terms of gas mass: even at a relatively modest
$\Sigma_{\rm HI} \approx 1$~M$_\odot$~pc$^{-2}$, one of our resolution
elements contains $\sim 5 \times 10^5$~M$_\odot$ of \hi. This is
enough material to make roughly 10 low-mass star-forming regions the
mass of the Taurus molecular complex --- perhaps a reasonable analog
to outer disk star forming regions given the low local SFRs and
sparsity of high mass stars. Detailed GALEX studies of outer disks
support this picture, finding star formation to be patchy, composed of
locally confined `FUV knots' \citep{gildepaz07a,thilker07} that may
match theoretical expectations for this regime
\citep[e.g.,][]{elmegreen06}. 

An even simpler explanation can be obtained by realizing that 
local conditions clearly influence the amount and efficiency of star
formation in systematic, measurable ways. However, the convolution of
a turbulent ISM, random sampling of star-forming clouds along their
evolutionary sequence, and the sensitivity of star formation tracers
to a range of stellar ages means that star formation `thresholds'
will always be observed as continuous trends. The physics discussed by
\citet{schaye04}, \cite{elmegreen94}, and \citet{kennicutt89} may
drive the trends that we observe, but we expect it is unlikely that
any proposed threshold will yield a clear step function on kpc scales
(and measurements on much smaller scales risk returning the trivial
result that stars form in dense, bound clouds rather than assess where
stars can and cannot form on galactic scales).

\subsection{What Regulates Star Formation In Outer Disks?}

We have seen outer disks to be distinct from the inner parts of
spirals in several ways: SFEs are low, the gradient in SFE with radius
is comparatively weak, and there is a clear spatial correlation between \hi\ and SFE.
We have also found the outer disks of dwarf and spiral galaxies to be similar in
many ways. All of these, we argue, point to \hi\ column as the
regulating quantity for star formation in outer disks. There are two
natural reasons for this: first, \hi\ represents the ultimate fuel for
star formation and unlike in inner galaxy disks the availability of
\hi\ varies dramatically across outer galaxy disks. Second, inasmuch
as the outer parts of galaxies are actually organized into disks, the \hi\ column
will usually represent the dominant (baryonic) mass component, meaning that the
{\em volume} density of gas (which depends critically on the stellar
surface density inside $r_{25}$) depends mostly on $\Sigma_{\rm
HI}$. Typical rank correlation coefficients
across the outer disks of our galaxy sample show a comparable degree of
local correlation ($r\approx0.4$) between the SFE and \hi\ column and between SFE and radius.
This can be (at least partly) understood as both quantities are strongly anti-correlated
(i.e., are not independent; compare Figure \ref{fig2}).
The correlation coefficient between SFE and Toomre-Q ($r\lesssim0.1$) implies little or no
local correlation between these quantities.

To first order a line of sight with a given \hi\ column appears to
form stars in a way that is largely independent of whether it lies in
the outer disk of a dwarf or spiral galaxy. There are second order
differences that point to the importance of other factors,
however. The most straightforward of these is the large scatter in
$\Sigma_{\rm SFR}$ at a given \hi\ column in spirals. Our estimate of
the intrinsic scatter in $\Sigma_{\rm SFR}$  near
$\Sigma_{HI} \sim 1$~M$_\odot$~pc$^{-2}$ is $\approx 1$~dex. Some of
this might be expected from the low values involved: the median SFR
for the lowest bins corresponds to forming only a few hundred solar
masses every 100 Myr. However, over the same range dwarf galaxies show
significantly less scatter; with such a small sample of dwarfs it is
hard to know whether the difference is significant, but if it is then
an easy interpretation is that the larger scatter in spirals reflects
the wider range of environments found there (e.g., our sample includes
spirals with strong metallicity gradients like M~101 and such with shallow
gradients like M~51 and M~83).

What can we say about the efficiency of the regulation of star formation in outer disks?
In the inner parts of galaxies stars form out of molecular clouds
with a depletion time of a few times $10^9$~yr. Across both dwarfs and
spirals we find \hi\ depletion times in outer galaxies
two orders of magnitude higher, $\sim 10^{11}$~yr. If
an outer disk star-forming cloud resembles its inner disk counterpart
(a big assumption), then only $\sim 1\%$ of the gas in outer galaxy disks
is actually in star forming clouds (or about $\sim 5,000$~M$_\odot$ in
one of our typical resolution elements). Similarly, if we take the
orbital time to be $\sim 0.5$~Gyr (roughly appropriate for
the Milky Way around $r_{25}$) then only $\sim 0.1\%$ of the gas is
converted into stars per orbital time, two orders of magnitude below the
galaxy-averaged value of \citet{kennicutt98}. Assuming a typical scale
height of $\sim 500$~pc, average volume densities in the outer disks
will be $\sim 0.1$~cm$^{-3}$, implying a disk free-fall time of $\sim
10^8$~yr. \citet{krumholz05} and \citet{krumholz07} argued that
star-forming structures on many scales convert $\sim 1\%$ of their gas
to stars per free-fall time. Outer galaxy disks apparently have about
a tenth that efficiency on average.

\subsection{Towards a Complete Star Formation `Law'}

Finally, we comment on the apparently `baroque' nature of the
`star formation law' plot seen in Figure \ref{fig11}. The plot
shows a complex star formation law, with at least three (somewhat) distinct regimes observed ---
outer disks, the \hi-dominated parts of inner disks, and the \htwo-dominated
parts of inner disks. Another regime is well-established:
starburst galaxies are known to have a steeper power law index than
the \htwo - dominated parts of spirals in Figure \ref{fig11}, so that
the plot should `turn up' again at $\Sigma_{\rm gas} \approx
100$~M$_\odot$~pc$^{-2}$ \citep[compare discussion in][]{bigiel08}.
Different parts of the `S-shape' (a `W-shape'
with the starbursts) represent physically truly distinct
regimes. We argue that this kind of behavior --- a
series of regime-dependent relations that combine into a complex
distribution --- is actually exactly
what one would expect if the ability to form stars or star-forming
clouds has a significant environmental dependence and multiple
environments are combined on a single plot.

\section{Summary}
\label{summary}

Using high resolution ($\sim 15\arcsec$), sensitive, wide
field-of-view data we have studied the relationship between \hi\ gas
and star formation traced by UV light in the outer disks
($1$--$2\times r_{25}$) of 17 spiral and 5 dwarf galaxies. We expect that
\hi\ dominates the ISM in this regime and that FUV reflects the
distribution of recently formed stars without large biases due to
internal extinction.

We find that:

\begin{enumerate}

\item Despite widespread star formation the gas depletion time is
  very low in outer galaxy disks, almost always longer than
  $10^{10}$~yr and typically $\sim 10^{11}$~yr. Star formation at the present
  rate will thus not appreciably deplete the gas supply in the outer
  disks over timescales shorter than many Hubble times.

\item FUV emission in outer disks declines approximately exponentially
  with a scale length comparable to that found for star formation in the inner
  parts of galaxies. \hi\ gas also usually shows a systematic decline, but a
  shallower one than the FUV. The \hi\ scale length that we derive for
  outer disks is typically twice the FUV scale length. As a result,
  the ratio of \hi\ to FUV and thus the gas (\hi) depletion time,
  increases with radius. However, this increase is weaker than that
  observed inside $r_{25}$, so that overall the depletion time as a
  function of radius becomes `flatter' as one moves out in a galaxy.

\item Unlike in the inner parts of galaxies, there is a fairly clear
  relationship between star formation and \hi\ in outer galaxy
  disks. This is clearest near the optical radius, where the \hi\ and
  FUV are strongly correlated (rank correlation $\sim
  0.7$). Extinction and the presence of H$_2$ wash out this correlation at radii $< r_{25}$  and the
  strength of the correlation also decreases with increasing radius ($r>r_{25}$), perhaps due
  to the increasing scatter/stochasticity in star formation at
  these radii. A fit of depletion time vs. $\Sigma_{\rm HI}$ to binned
  outer disk data yields $\tau_{\rm Dep} \propto (\Sigma_{\rm
    HI})^{0.7}$, or $\Sigma_{\rm SFR} \propto (\Sigma_{\rm HI})^{1.7}$
  for both spiral and dwarf galaxies.

\item At a given \hi\ column density, one finds similar FUV
  intensities in the outer disks of dwarf and spiral galaxies. The
  main difference between the two samples is the distribution of
  \hi\ column densities at large radius. Dwarf galaxies tend to have
  (relatively) more high-column \hi\ than spirals. In both samples, low columns
  become more common with increasing radius.

\item Examining the relationship for individual lines of sight, we do
  not find clear evidence for a step-function type star formation
  threshold in either radius or \hi\ column, though the SFE is a weak
  function of both quantities.

\item The relationship between $\Sigma_{\rm SFR}$ and $\Sigma_{\rm
  gas}$ for individual lines of sight extends that measured by
  \citet{bigiel08} into a new regime of low $\Sigma_{\rm SFR}$ and low
  $\Sigma_{\rm gas}$ more or less continuously. In this regime, a
  line of fixed $\tau_{\rm Dep} \sim 10^{10}$~yr forms an approximate
  upper envelope to the distribution. Lines of sight with high \hi\ columns ($\Sigma_{\rm gas}
  \approx 3$--$10$~M$_\odot$~pc$^{-2}$) but little or no FUV emission are
  rare. These conclusions apply to both dwarf and spiral galaxies.
  
\end{enumerate}

\acknowledgments We thank the GALEX NGS team for creating and making available
their outstanding dataset. F.B. acknowledges support from NSF grant AST-0838258
and earlier support from the Deutsche Forschungsgemeinschaft (DFG) Priority Program 1177.
Support for A.L. was provided by NASA through Hubble Fellowship grant
HST-HF-51258.01-A awarded by the Space Telescope Science Institute, which
is operated by the Association of Universities for Research in Astronomy,
Inc., for NASA, under contract NAS 5-26555. The work of W.J.G.d.B. 
is based upon research supported by the South African Research Chairs Initiative
of the Department of Science and Technology and National Research Foundation. 
We have made use of the Extragalactic Database (NED), which is operated by the Jet
Propulsion Laboratory, California Institute of Technology, under
contract with the National Aeronautics and Space Administration. This
research has made use of NASA's Astrophysics Data System (ADS).

\end{document}